\documentclass[a4paper]{article}

\title{A Time Series Approach to Explainability for Neural Nets with Applications to Risk-Management and Fraud Detection}
\author{Marc Wildi\footnote{ZHAW Zurich University of Applied Sciences, IDP; e-mail: wlmr@zhaw.ch} and Branka Hadji Misheva\footnote{BFH, Bern University of Applied Sciences, Institute for Applied Data Science and Finance; email: heb1@bfh.ch}}

\usepackage{a4wide}
\usepackage{amsmath}
\usepackage[utf8]{inputenc}
\usepackage{hyperref}
\usepackage{float}

\usepackage{Sweave}
\begin{document}

\maketitle

\begin{abstract}

Artificial intelligence (AI) is creating one of the biggest revolution across technology-driven application fields. For the finance sector, it offers many opportunities for significant market innovation and yet broad adoption of AI systems heavily relies on our trust in their outputs. Trust in technology is enabled by understanding the rationale behind the predictions made. To this end, the concept of eXplainable AI (XAI) emerged introducing a suite of techniques attempting to explain to users how complex models arrived at a certain decision. For cross-sectional data classical XAI approaches can lead to valuable insights about the models' inner workings, but these techniques generally cannot cope well with longitudinal data (time series) in the presence of dependence structure and non-stationarity. We here propose a novel XAI-technique for deep learning methods (DL) which preserves and exploits the natural time ordering of the data. Simple applications to financial data illustrate the potential of the new approach in the context of risk-management and fraud-detection.

\end{abstract}

\section{Introduction}
Developing accurate forecasting methodologies for financial time series remains one of the key research topics relevant from both a theoretical and applied viewpoint. Traditionally, researchers aimed at constructing a causal model, based on econometric modelling, that explains the variations in the specific time series as a function of other inputs. Yet, traditional approaches often struggle when it comes to modelling high-dimensional, non-linear landscapes often characterized with missing or sparse input space.\\

Recently, deep learning (DL) has become highly popularized in many aspects of data science and has become increasingly applied to forecasting financial and economic time series (\cite{article1}, \cite{Persio2017RecurrentNN}, \cite{unknown1}, \cite{yang2021forecasting}). Recurrent methods are suited to time series modelling due to their memory state and their ability to learn relations through time; moreover, convolutional neural networks (CNN) are also able to build temporal relationships (\cite{rojat2021explainable}). The literature offers various examples of the application of DL methods to stock and forex market forecasting, with results that significantly outperform traditional counterparts (\cite{article2},\cite{Deniz}, \cite{inproceedings}, \cite{Zexin}). This was also confirmed in more recent installments of the Makridakis Forecasting Competitions which have been held roughly once a decade since 1982 and have the objective of comparing the accuracy of different forecasting methods. A recurring conclusion from these competitions has been that traditional, simpler methods are often able to perform equally well as their more complex counterparts. This changed at the latest editions of the competition series, M4 and M5, where a hybrid Exponential Smoothing Recurrent Neural Network method and LightGBM, won the competitions, respectively (\cite{MAKRIDAKIS202054};\cite{MAKRIDAKIS2021}).\\

The introduction of DL methods for financial time series forecasts potentially enables higher predictive accuracy but this comes at the cost of higher complexity and thus lower interpretability. DL models are referred to as “black boxes” because it is often difficult to understand how variables are jointly related to arrive at a certain output. This reduced ability to understand the inner workings and mechanisms of DL models unavoidably affects their trustworthiness and the willingness among practitioners to deploy such methods in sensitive domains such as finance. As a result, the scientific interest in the field of eXplainable artificial intelligence (XAI) had grown tremendously within the last few year (\cite{e23010018}, \cite{Dazeley}, \cite{Gunning}). XAI aims at introducing a suite of techniques attempting to communicate understandable information about how an already developed model produces its predictions for any given input (\cite{arrieta2019explainable}). In terms of the taxonomy of XAI methods, the literature offers a comprehensive overview of the existing research in the field of XAI (see \cite{arrieta2019explainable}, \cite{Maksymiuk}, \cite{Samek}). In general, methods are considered in view of four main criteria (\cite{Linardatos}): (i) the type of algorithm on which they can be applied (model-specific vs model-agnostic), (ii) the unit being explained (if the method provides an explanation which is instance-specific then this is a local explainability technique and if the method attempts to explains the behavior of the entire model, then this is a global explainability technique), (iii) the data types (tabular vs text vs images), and (iv) the purpose of explainability (ex. improve model performance, test sensitivity, etc.).
The growing popularity of the topic notwithstanding, research on XAI in finance remains limited and most of the existing explainability techniques are not suited for time series, let alone for non-stationary financial time series and their somehow notorious stylized facts. Many state-of-the-art XAI methods are originally tailored for certain input types such as images (ex. Saliency Maps) or text (ex. LRP) and have later been adjusted to suit tabular data as well. However, the temporal dimension is often omitted and the literature currently offers only a limited consideration of the topic. Notable examples are interpretable decision trees for time series classification (\cite{Hidasi}) and using attention mechanisms (\cite{Hsu} and \cite{schockaert2020attention}), with none of the applications looking specifically at explainability for financial data. \\

To address this gap in the literature as well as in the considered application field, we propose a generic XAI-approach for neural nets based on a family of X-functions which preserve and exploit the natural time ordering and the possible non-stationary dependence structure of the data. We here propose a set of potentially interesting X-functions pertinent for a broad range of financial applications and we derive explicit formula for two specific family members which address (non-)linearity of the model and by extension (non-)linearity and (non-)stationarity of the data and the underlying data generating process. Our empirical examples, based on applications of these X-functions, suggest evidence of a perceptual hierarchy of 'explanations': on a macro-level, net-responses appear to be determined mainly by random effects imputable to imperfect numerical optimization and random initialization of net-parameters; moreover, the extent of these random effects is for the most part unaffected by the choice of optimization algorithm, software implementation (package) or complexity of the net (architecture and number of neurons); finally and not least surprisingly, once the random effects are sorted out, the richly parameterized non-linear nets resemble  well-known  forecast heuristics whose unassuming simplicity eludes the need for explainability. On a micro-level, the proposed X-functions expose changes in the data generating process of the series by displaying unusual input-output relations marking episodes of higher uncertainty or 'unusual' market activity.  We then argue that simple statistical techniques applied to the X-functions' data-flow, instead of the original time series, could provide potential added-value in a broad range of application fields, including risk-management and fraud detection for example.\\

Before entering the proper topic, let us reaffirm that the lack of explainability represents currently one of the most relevant barriers for wider adoption of DL in finance. This challenge has become particularly relevant for European finance service providers as they are subjected to the General Data Protection Regulation (GDPR) which provides a right to explanation, enabling users to ask for an explanation as to an automated decision-making process affecting them. By proposing explainability techniques suited to the context of financial time series, we enable practitioners \textit{"to have the cake and eat it too"} - i.e to utilize both the predictive accuracy of DL methods while at the same time maintaining a sufficient level of explainability as to the predictions obtained. \\

\section{Explainability of Neural Nets: a Selection of Classic Approaches}

\subsection{Classical Approaches: An Overview}
The literature offers different viewpoints concerning the classification of the many emerging interpretability methods. A particular relevant distinction among interpretable models is based on whether the specific technique is model-specific or model agnostic. Among the explainability techniques that are specifically tailored to DL models, substantial portion of researchers' attention is focused on applications featuring image data (ex. saliency map (\cite{Itti2009AMO}), gradients (\cite{simonyan2014deep}), deconvolution (\cite{zeiler2013visualizing}), class activation map (\cite{zhou2015learning})). The model-agnostic methods, on the other hand, are being increasingly applied to explain black-box models fitted to financial data (\cite{Giudici1}, \cite{Niklas}, \cite{misheva2021explainable}). Global, model-agnostic methods include, among others, the partial dependency plot (PDP), the accumulated local effects (ALE) and the permutation feature importance (PFI). PDPs help visualize the relationship between a subset of the features and the response while accounting for the average effect of the other predictors in the model. For numerical data, the PD-based feature importance is defined as a deviation of each unique feature value from the average curve (\cite{molnar2019}):
\[I_{(x_s)}=\sqrt{\frac{1}{K-1}\sum_{k=1}^k \left(f_s({x}_s^{k}) - \frac{1}{K}\sum_{k=1}^k f_s({x}_s^{k})\right)^2}\]
where ${x}_s^{k}$ are the K unique values of the variable $\mathbf{x_s}$. A Key assumption of the PDPs is feature independence i.e. the approach assumes that the variables for which the partial dependence is computed are not correlated with other features. ALE plots emerged as an unbiased alternative of the PDPs. Differently from the PDP, ALE can deal with feature correlations because they average and accumulate the difference in predictions across the conditional distribution, which isolates the effects of the specific feature. Formally (\cite{molnar2019}):
\[f(x_{[i]})=\frac{1}{T}\sum_{t=1}^T NN(x_{[i]},\mathbf{x}_t^{-i})\]
\begin{itemize}
\item The $t$-th observation is $\mathbf{x}_t=(x_{1t},...,x_{mt})$
\item $\mathbf{x}_t^{-i}$ is the $t$-th observation without the i-th explanatory (its dimension is $m-1$)
\item $x_{[i]}$ is the $i$-th explanatory arranged from the smallest to the largest observation
\item $f(x_{[i]})$ is the mean-output of the net when keeping $x_{[i]}$ fixed and sampling over all $\mathbf{x}_t^{-i}$
\end{itemize}
PFI is yet another global, model-agnostic interpretability technique based on the work of \cite{fisher2019models}. PFI measures the increase in the model's prediction error after feature permutation is performed. Under this approach a feature is considered irrelevant if after its permutation, the model's error remains unchanged. \\

Among the model-agnostic techniques, which explain individual predictions or classifications, two frameworks have been widely recognized as the state-of-the-art methods and those are: (i) the LIME framework, introduced by \cite{ribeiro2016why} in 2016 and (ii) SHAP values, introduced by \cite{lundberg2017unified}. LIME, short for locally interpretable model agnostic explanations, is an explanation technique which aims to identify an interpretable model over the representation of the data that is locally faithful to the classifier \cite{ribeiro2016why}. Specifically, LIME disregards the global view of the dependence between the input-output pairs and instead derives a local, interpretable model using sample data points that are in proximity of the instance to be explained. For further details, see \cite{ribeiro2016why}. More formally, the explanations provided by LIME is obtained by following (\cite{molnar2019}):
\begin{equation}
\xi (x)=  argmin_{g\in G} \ L(f, g, \pi _{x}) + \Omega (g)
\end{equation}
where,

$\emph g$ : An explanation considered as a model

$\emph g \in G$: class of potentially interpretable models such as linear models and  decision trees

$\mathcal{\emph f: R^ d \rightarrow R}$: The main classifier being explained

$\mathcal{\pi_x (z)}$: Proximity measure of an instance $\emph z$ from x

$\mathcal{\Omega (\emph g)}$: A measure of complexity of the  explanation $g \in G$\\

The goal is to minimize the locality aware loss \emph{L} without making  any assumptions about \emph{f}. SHAP, short for SHapley Additive exPlanations, presents a unified framework for interpreting predictions \cite{lundberg2017unified}. According to the paper by \cite{lundberg2017unified}, for each prediction instance, SHAP assigns an importance score for each feature included in the model's specification. Its novel components include: (i) the identification of a new class of additive feature importance measures, and (ii) theoretical results showing there is a unique solution in this class with a set of desirable properties.

\subsection{The Utility of Classical Approaches for Applications Featuring Financial Time Series}

Classical approaches and their current implementation are not tailored for financial data hence their applicability in this domain is limited. Key limitation of many classical methods is the fact that they ignore feature dependence which is a defining property of financial data. Specifically, the procedures of perturbation-based methods like the PDP, PFI, SHAP, LIME, etc., start with producing artificial data points, obtained either through replacement with permuted or randomly select values from the background data; or through the generation of new "fake" data, that are consequently used for model predictions. Such step results in several concerns:
\begin{itemize}
    \item if features are correlated, the artificial coalitions created will lie outside of the multivariate joint distribution of the data,
    \item if the data are independent, coalitions can still be meaningless; perturbation-based methods are fully dependent on the ability to perturb samples in a meaningful way which is not always the case with financial data (ex. one-hot encoding)
    \item generating artificial data points through random replacement disregards the time sequence hence producing unrealistic values for the feature of interest.
\end{itemize}
In any case and notwithstanding the possibility of mixing  unrelated trend-levels and volatility-clusters, the eventuality of re-combining artificially data from remote past and current time seems counter-intuitive not least from a purely application-based 'meta-explainability' perspective. Yet another concern with classical approaches is that they can lead to misleading results due to feature interaction. Namely, as demonstrated by (\cite{molnar2021general}) both PDP and ALE plot can lead to misleading conclusions in situations in which features interact.

Looking specifically at the SHAP framework, both conditional and marginal distribution can be used to sample the absent features and both approaches have their own issues. For example, the TreeSHAP is a conditional method and under conditional expectations a feature that has no influence on the prediction function (but is correlated with another feature that does) can get a TreeSHAP estimate different from zero (\cite{molnar2019}) and can affect the importance of the other features. For examples featuring real data sets see \cite{chen2020true}.

On the other hand, sampling from the marginal distribution, instead of the conditional, would ignore the dependence structure between present and absent features. Consider an example in which we build a model to predict prices of apartments and we have some variables that describe the apartments as inputs (ex. size of the apartment, location, number of rooms, number of bathrooms, etc.). Let's further assume that in one coalition the size of apartment feature is equal to 24 square meters and we sample values for the number of rooms feature (i.e. the absent feature). If we sample a value of 6 for this absent feature, we have created an unrealistic data point that lies off the true data manifold and it is further used to evaluate the model. Further discussion and examples on the mathematical issues that arise from the estimation procedures used when applying Shapley values as feature importance measures can be found in \cite{kumar2020problems}.


\section{XAI for Neural Nets: a Time Series Approach }\label{xai_nn}

Until recently, computationally intensive methods had a hard time competing against classic (linear) time series approaches, at least in the context of large-scale international forecast competitions, see \cite{cronehibbon2011} for a review\footnote{The author of this footnote won the NN3 and NN5 forecast competitions, documented in the referenced article, by relying on a linear approach.}. However, the recent M4 and M5 competitions brought forward hybrid approaches mixing neural nets and classic exponential smoothing, see \cite{SlawekSmyl2020}. These encouraging results pushed us to proceed to a more comprehensive analysis of neural-nets, as applied to time series forecasting, by addressing the notorious opacity or 'black-box' problem in a way compliant with longitudinal or cross-sectional dependency. 
In order to introduce the topic we first point towards an identification problem which impedes an interpretation of the actual net-parameters.

\subsection{Random Nets and Indeterminacy of Ordinary Net Parameters}\label{r_toy}

\begin{figure}[H]\begin{center}\includegraphics[height=3in, width=3in]{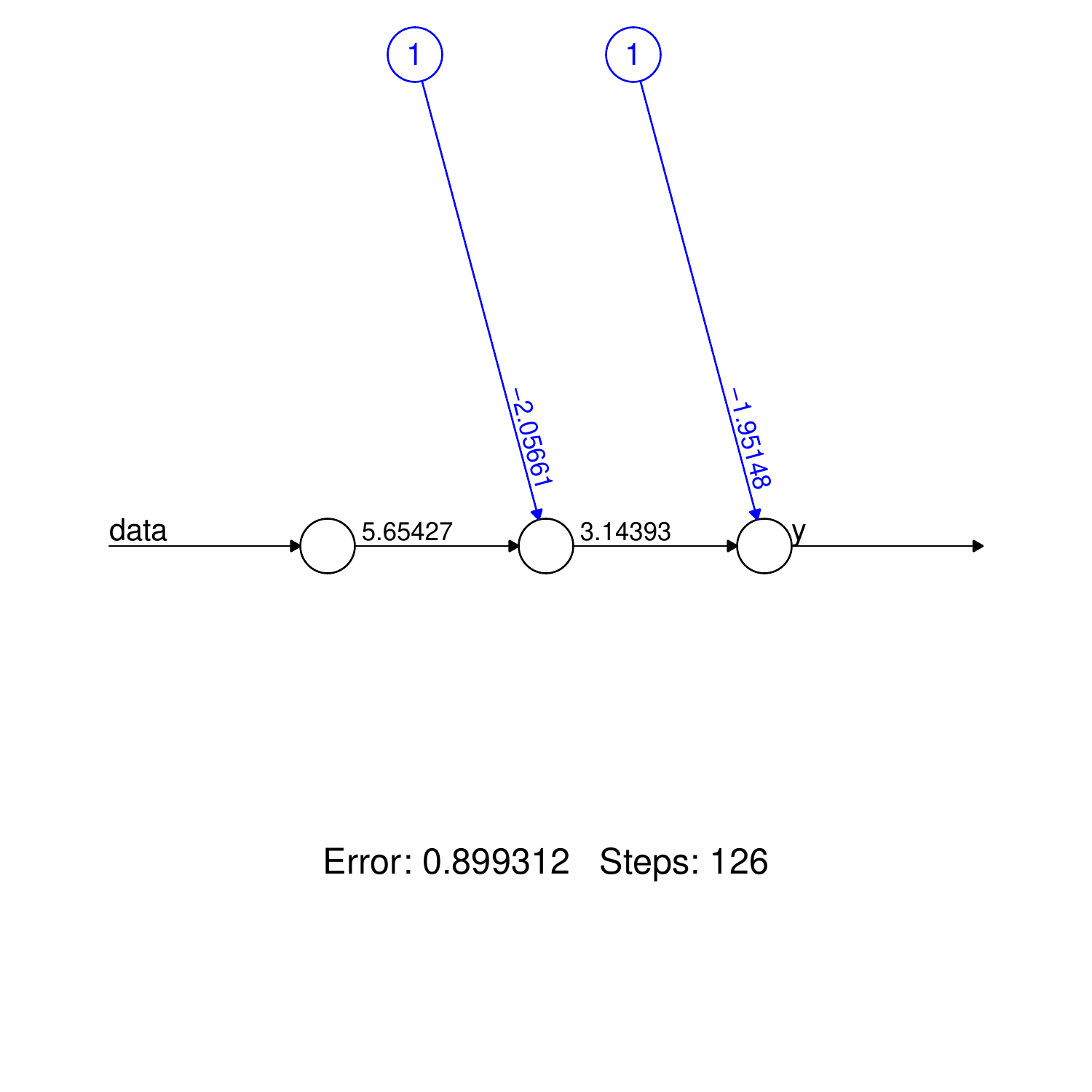}\caption{Toy net: estimation based on neuralnet-package with set.seed(1)\label{nn_toy1}}\end{center}\end{figure}

For illustration we here rely on a minimalist single-neuron single hidden-layer feedforward architecture, a 'toy-net', as shown in fig.\ref{nn_toy1}.
This net is applied to artificial data generated according to the following model
\[y_t=x_t+\epsilon_t\]
where $x_t,\epsilon_t$ are independent realizations of standard Gaussian noise: the toy-net must learn the model by fitting $y_t$ (target) based on $x_t$ (net input). Since we opted for the classic sigmoid activation function  $\sigma(x)=\frac{1}{1+exp(-x)}$, the data is previously maped into the unit-interval $x_t':=\frac{x_t-\min(x_t)}{\max(x_t)-\min(x_t)},y_t':=\frac{y_t-\min(y_t)}{\max(y_t)-\min(y_t)}$ so that the ranges of net-output and of target match (the transformation of the input $x_t$ is less relevant in this example). For better interpretation all reported mean-square (MSE-) performances refer to back-transformed   original data $x_t,y_t$. 
The following non-linear function is obtained
\[o_t=\sigma\Big(-1.951+3.14~\sigma(-2.057+5.65~x_t)\Big)\]
after fitting the net to the data, where $o_t$ designates the net-output and where biases $b_1=-2.057,b_2=-1.951$ (blue numbers in fig.\ref{nn_toy1}) and weights $w_1=5.654,w_2=3.144$ (black numbers in fig.\ref{nn_toy1}) are obtained by applying classic \emph{backpropagation}  as implemented in the \emph{neuralnet} package of the \textbf{R} software environment, assuming a particular random-initialization for the unknown parameters. Surprisingly, different solutions are obtained depending on the random initialization of the parameters and markedly different estimates are obtained when relying on our own steepest-gradient algorithm, see table \ref{parm_ind}.
\begin{table}[ht]
\centering
\begin{tabular}{rrrr}
  \hline
 & neuralnet seed(1) & neuralnet seed(6) & own steepest gradient seed(1) \\ 
  \hline
b1 & -2.057 & 0.703 & 2.122 \\ 
  w1 & 5.654 & -4.156 & 1.715 \\ 
  b2 & -1.951 & 1.255 & -40.445 \\ 
  w2 & 3.144 & -5.146 & 42.696 \\ 
  mse & 0.899 & 0.894 & 0.890 \\ 
   \hline
\end{tabular}
\caption{Estimated parameters and MSE-performances for different random-seeds and optimization algorithms} 
\label{parm_ind}
\end{table}
The 'final' criterion values in the last row reveal that the numerical optimization does not converge to the global optimum so that 'final' estimates are subject to an additional layer of randomness, inherited from parameter-initialization. Furthermore, the table suggests evidence of an identification problem since nearly identical criterion values are obtained  based on wildly different estimates. In the sequel we will refer to these problems by the term \textrm{random net} to signify that a particular realization or instance of a neural net is dependent on the initialization of its parameters. Obviously, under these circumstances, a proper explanation of the net based solely on its weights or biases seems compromised, particularly for increasingly complex deep nets. 
Note also that alternative packages such as keras (tensorflow) or MXNet do not fix these issues, at all.\\


Based on these empirical evidences we now propose a way towards explainability which bypasses the inner structure of the net by emphasizing input-output relations instead, as ordered by time.

\subsection{Explainability: a Time Series Approach}\label{time series}

In order to preserve data-integrity as well as model-integrity we propose to analyze the effect of infinitesimal changes of the explanatory variables on some function of the net-output at each time-point $t=1,...,T$. Extensions to discrete-valued data, for example classes, is discussed below.

\subsubsection{X-Functions}

We here propose a family of potentially interesting explainability (X-)functions $xf(\cdot)$ for assigning meaning to the net's response or output $\mathbf{o_t}$ over time $t=1,...,T$, where $\mathbf{o_t}=(o_{1t},...,o_{n_pt})$ is a $n_p$ dimensional vector of output neurons. 
By selecting the identity $xf(\mathbf{o_t})=\mathbf{o_t}$ we can mark preference for the sensitivities or partial derivatives $w_{ijt}:=\partial o_{jt}/\partial x_{it}$, $i=1,...,n$, $j=1,...,n_p$, for each explanatory variable $x_{it}$ of the net. In order to complete the 'explanation' derived from the identity one can add a synthetic intercept to each output neuron $o_{jt}$ defined according to
\begin{eqnarray}\label{intercept}
b_{jt}:=o_{jt}-\sum_{i=1}^n w_{ijt}x_{it}
\end{eqnarray}
For each output neuron $o_{jt}$, the resulting derivatives or 'explanations' $b_{jt},w_{1jt},...,w_{njt}$ generate a new (heavily transformed) data-sample which is referred to as Linear Parameter Data or LPD for short: the LPD is a matrix of dimension $T*(n+1)$, irrespective of the complexity of the neural net, with $t-$th row denoted by $\mathbf{LPD}_{jt}:=(b_{jt},w_{1jt},...,w_{njt})$. The LPD can be interpreted in terms of exact replication of the net by a linear model at each time point $t$ and the \emph{natural} time-ordering of $\mathbf{LPD}_{jt}$ subsequently allows to examine changes of the linear replication as a function of time. We are then in a position to assign a meaning to the neural net, at each time point $t=1,...,T$, and to monitor non-linearities of the net or, by extension, possible non-stationarities of the data as illustrated in the empirical section. Further statistical analysis could be applied to the LPD in order to detect, assess or track e.g. unusual observations (outliers) or unusual dynamics (fraud). Specifically, we here suggest that an application of well-known time series techniques to the LPD, i.e. $\mathbf{LPD}_{jt}$, instead of the original data $x_{1t},...,x_{nt}$, may reveal new features in the data as measured by unusual sensitivity of the net-output to the net-input, see below for details.

\subsubsection{Derivation of Linear Parameter Data (LPD) and of Arbitrary Differentiable X-Functions}

We first derive a formal expression for the LPD, which corresponds to the special case when the X-function is the identity, based on forward- as well as backward-sequences
i.e. proceeding from right to left or from left to right along the chain-rule of differentiation
of the non-linear net-output $\mathbf{o_t}$: both expressions are required later when deriving corresponding
formal expressions for another explainability function, namely the Quadratic Parameter Data or
QPD for short. We first proceed by the forward-sequence and assume a (feedforward)
neural net with $p-1$ hidden layers, whereby the $k$-th hidden layer has dimension $n_k$,
corresponding to its number of neurons; we also assume that all neurons have a sigmoid activation
function (straightforward modifications apply in the case of arbitrary differentiable activation
functions). Let $\mathbf{A}^{(k)}$ denote a column-vector of dimension $n_k$ corresponding
to the vector of outputs of the $n_k$ neurons in the $k$-th layer at time $t$ and let $\mathbf{W}^k$ designate the matrix of dimension $(n_{k-1},n_k)$ of weights linking the neurons in layer $k-1$ to the neurons in layer $k$ in the fully-connected feedforward net, whereby $n_0=n$ is the dimension of the input-layer (if the net is not fully connected then silent connections receive value zero in $\mathbf{W}^k$). We can then relate the outputs at layers $k-1$ and $k$ by
\begin{eqnarray}\label{kkm1}
\mathbf{A}^{(1)}_t&=&\mathbf{\sigma}\left(\mathbf{W}^1~' \mathbf{x}_t\right)\\
\mathbf{A}^{(k)}_t&=&\mathbf{\sigma}\left(\mathbf{W}^k~' \mathbf{A}^{(k-1)}_t\right)\label{kkm2}
\end{eqnarray}
where $'$ (apostrophe) refers to the ordinary matrix transposition, $\mathbf{x}_t$ is the $n$-dimensional vector of input-data and where $\sigma()$ could be interpreted as a place-holder for any differentiable activation function, although we shall relate to the sigmoid function specifically when computing derivatives.  Denote further by $\mathbf{dA^f}^{(k)}_t$  the $(n,n_k)$-dimensional matrix of partial derivatives of the vector $\mathbf{A}^{(k)}$ with respect to the explanatory variables $x_{it},i=1,...,n$: the LPD is then identified with $\mathbf{dA^f}^{(p)}_t$ at the output neuron(s). Note that the superscript $\mathbf{f}$ in $\mathbf{dA^f}^{(k)}_t$
refers to the forward direction, computing the derivative from left (input-layer) to right (output-layer). 
For the first hidden layer, $k=1$, we then obtain the derivative of \ref{kkm1} as
\begin{eqnarray}\label{da_f1}
\mathbf{dA^f}_{t}^{(1)}=\Big(\mathbf{W}^{(1)}~'\cdot \mathbf{A}_t^{(1)}\cdot(\mathbf{e}-\mathbf{A}^{(1)}_t)\Big)'
\end{eqnarray}
where $\mathbf{e}$ is a column-vector of ones of dimension $n_1$ and where the multiplication symbol (dot) indicates element-wise multiplication i.e. $\mathbf{A}_t^{(1)}\cdot (\mathbf{e}-\mathbf{A}^{(1)}_t)$ is a column-vector of dimension $n_1$ obtained by multiplying the $i$-th element, $i=1,...,n_1$, of the column-vector $\mathbf{A}_t^{(1)}$ with the corresponding $i-$th element of $(\mathbf{e}-\mathbf{A}^{(1)}_t)$: the resulting vector corresponds to the derivative $\dot{\sigma}(\cdot)=\sigma(\cdot)(1-\sigma(\cdot))$ of the sigmoid in \ref{kkm1}\footnote{Straightforward modifications apply in the case of alternative activation functions.}. Similarly, the $j$-th row, $j=1,...,n_1$, of the transposed matrix $\mathbf{W}^{(1)}~'$ is multiplied with the $j-$th element of the vector $\mathbf{A}_t^{(1)}\cdot(\mathbf{e}-\mathbf{A}^{(1)}_t)$. As a result, the derivative $\mathbf{dA^f}_{t}^{(1)}$ is a matrix of dimension $(n,n_1)$. Having all necessary algebraic elements in place, we can now proceed iteratively for $k=2,...,p$, through all layers, obtaining the derivative of \ref{kkm2} as
\begin{eqnarray}\label{da_f2}
\mathbf{dA^f}_{t}^{(k)}=\Bigg(\Big(\mathbf{W}^{(k)}~'\mathbf{dA^f}_t^{(k-1)}~'\Big)\cdot \mathbf{A}_t^{(k)}\cdot(\mathbf{e}-\mathbf{A}^{(k)}_t)\Bigg)'
\end{eqnarray}
where $\mathbf{W}^{(k)}~'\mathbf{dA^f}_t^{(k-1)}~'$ is the ordinary matrix-product of the $(n_{k},n_{k-1})$-dim $\mathbf{W}^{(k)}~'$ and the $(n_{k-1},n)$-dim $\mathbf{dA^f}_t^{(k-1)}~'$ thus resulting in a $(n,n_k)$-dim matrix of derivatives $\mathbf{dA^f}_{t}^{(k)}$ in \ref{da_f2}, after transposition. If the output-layer consists of a single neuron, as assumed in our examples below, the forward-propagation $\mathbf{dA^f}_{t}^{(p)}$ corresponds to the ordinary gradient $\partial o_t/\partial x_{it}, i=1,...,n$ or $\mathbf{LPD}_t$ (without the artificial intercept defined by \ref{intercept} which has to be computed separately and added to the LPD); otherwise, $\mathbf{dA^f}_{t}^{(p)}$ corresponds to the Jacobi-matrix of partial derivatives for all output neurons. Note that this expression is time-dependent and that it differs from the time-invariant gradients with respect to net-parameters (weights and biases), as computed by the ordinary backpropagation-algorithm. \\
We now proceed to the alternative backward-sequence, determining the LPD in inverse direction along the chain-rule decomposition, and starting at the output layer
\begin{eqnarray}\label{da_b1}
\mathbf{dA^b}_{t}^{(p)}=\mathbf{A}_t^{(p)}\cdot(\mathbf{e}-\mathbf{A}^{(p)}_t)
\end{eqnarray}
where the superscript \textbf{b} in $\mathbf{dA^b}^{(p)}_t$ refers to the backward direction. Then recursively for $k=p-1,p-2,...,1$
\begin{eqnarray}\label{da_b2}
\mathbf{dA^b}_{t}^{(k)}=\Big(\mathbf{W}^{(k+1)}\cdot \mathbf{A}_t^{(k)}\cdot(\mathbf{e}-\mathbf{A}^{(k)}_t)\Big)\mathbf{dA^b}_t^{(k+1)}
\end{eqnarray}
where $\mathbf{W}^{(k+1)}$ has dimension $(n_k,n_{k+1})$ and the column-vector $\mathbf{dA^b}_t^{(k+1)}$ has dimension $n_{k+1}$ such that $\mathbf{dA^b}_{t}^{(k)}$ has dimension $n_k$, as required. Finally, for the input layer $k=0$ we obtain
\begin{eqnarray}\label{da_b3}
\mathbf{dA^b}_{t}^{(0)}=\mathbf{W}^{(1)}\mathbf{dA^b}_t^{(1)}
\end{eqnarray}
where $\mathbf{A}_t^{(0)}\cdot(\mathbf{e}-\mathbf{A}^{(0)}_t)$ is replaced by the identity (because input neurons do not have an activation function) and where $\mathbf{dA^b}_{t}^{(0)}$ is of dimension $n$: this expression  corresponds to an alternative derivation of the LPD at time $t$, at least up to the synthetic intercept specified by \ref{intercept}. Note that in the case of multiple output neurons ($n_p>1$) the column-vectors $\mathbf{dA^b}_{t}^{(k)}$ of dimension $n_k$ become matrices of dimension $(n_k,n_p)$ and $\mathbf{dA^b}_{t}^{(0)}$ is once again the Jacobi-matrix of dimension $(n,n_p)$ collecting all partial derivatives of the output neurons.\\

To conclude, sensitivities or partial derivatives of arbitrary differentiable X-functions $xf(\mathbf{o_t})$, where $\mathbf{o_t}=\mathbf{A}_t^{(p)}$, can be obtained straightforwardly from the above derivation of the LPD, by
substituting the composite activation function $xf(\sigma(\cdot))$ to $\sigma(\cdot)$ at the output neurons.  Specifically, for $k=p$, equation \ref{da_f2} is then augmented with the $n_p$-dimensional gradient $\mathbf{\nabla xf}(\mathbf{A}_t^{(p)})$ of the X-function $xf(\cdot)$ (with respect to the output-neurons) i.e.
\begin{eqnarray*}
\mathbf{d(xf\circ A^f)}_{t}^{(p)}:=\left(\mathbf{\nabla xf}(\mathbf{A}_t^{(p)})\cdot\mathbf{dA^f}_{t}^{(p)}~'\right)'
\end{eqnarray*}
where the $(n,n_p)$-dimensional $\mathbf{d(xf\circ A^f)}_{t}^{(p)}$ denotes the Jacobi-matrix of the sought-after sensitivities and where the $n_p$-dimensional (column-) vector $\mathbf{\nabla xf}(\mathbf{A}_t^{(p)})$ is the ordinary gradient $\partial xf(\mathbf{o_t})/\partial o_{it}$, $i=1,...,n_p$, of $xf(\cdot)$ computed at the output layer. A similar extension applies to \ref{da_b1}
\begin{eqnarray}\label{da_b1xf}
\mathbf{d(xf\circ A^b)}_{t}^{(p)}:=\mathbf{\nabla xf}(\mathbf{A}_t^{(p)})\cdot\mathbf{dA^b}_{t}^{(p)}
\end{eqnarray}
where now all involved terms are $n_p$-dimensional column vectors.

\subsubsection{Derivation of Quadratic Parameter Data (QPD) and of Arbitrary Twice-Differentiable X-Functions}

Another potentially interesting explainability function is obtained when identifying $xf(\cdot)$ with the previous LPD so that $xf(o_{jt})=\mathbf{LPD}_{jt}=\partial o_{jt}/\partial x_{it}$, $i=1,...,n$, $j=1,...,n_p$: the derived sensitivity, in terms of second order partial derivatives, $\frac{\partial^2 o_{jt}}{\partial x_{it}\partial x_{kt}}$ defines a new data-sample referred to as Quadratic Parameter Data or QPD. The QPD is a measure of change of the LPD at each time point and can be interpreted as a measure for non-linearity of the net or, by extension, for non-linearity or non-stationarity of the data generating process, along the time axis. While for each neuron $o_{jt}$ the new data-sample generated by the QPD is a three-dimensional array of dimension $T*n*n$, we are often mainly interested in the diagonal elements $\frac{\partial^2 o_{jt}}{\partial^2x_{it}}, i=1,...,n$, so that the corresponding diagonal flow has dimension $T*n$, the same as the LPD. For each output-neuron $o_{jt}$, $j=1,...,n_p$, a formal expression of the QPD can be obtained by differentiating the backward-equations \ref{da_b1}, \ref{da_b2} and \ref{da_b3}, breaking-up the chain-rule of first-order differentiation into forward branch, for the inner functions, and backward branch, for the outer function. Specifically, starting at the output layer i.e. at $o_{jt}$ we obtain\footnote{We here focus on single neurons in order to avoid the appearance of cumbersome three-dimensional arrays.}
\begin{eqnarray}
{dda}_{jt}^{(p)}&=&{A}_{jt}^{(p)}(1-{A}^{(p)}_{jt})(1-2{A}^{(p)}_{jt})\label{dda}\\
\mathbf{ddA}_{jt}^{(p)}&=&{dda}_{jt}^{(p)}\Big(\mathbf{W}^{(p)}_j~'\mathbf{dA^f}_{t}^{p-1}~'\Big)'\label{ddA}
\end{eqnarray}
where \ref{dda} corresponds to the second order derivative of the sigmoid $\ddot{\sigma}()=\sigma()(1-\sigma())(1-2\sigma())$, $\mathbf{dA(f)}_t^{p-1}$ in \ref{ddA} is the forward-derivative obtained in 
\ref{da_f2} and $\mathbf{W}^{(p)}_j$  is the $j$-th column of $\mathbf{W}^{(p)}$, of dimension $n_{p-1}$, linking the neurons in layer $p-1$ to the $j$-th output neuron.  Note that $\mathbf{dA^f}_{t}^{p-1}$ has dimension $n,n_{p-1}$ so that $\mathbf{dA^f}_{t}^{p-1}\mathbf{W}^{(p)}_j$ and hence $\mathbf{ddA}_{jt}^{(p)}$ have dimension $n$. As claimed, \ref{ddA} is the derivative of \ref{da_b1} at the $j$-th output neuron whereby ${dda}_{jt}^{(p)}$ is the derivative of the outer function and the forward-term $\mathbf{W}^{(p)}_j~'\mathbf{dA^f}_{t}^{p-1}~'$ is the derivative of the inner function(s). We can now iterate backwards through all additional hidden layers $k=p-1,p-2,...,2$ according to
\begin{eqnarray}
\mathbf{dda}_t^{(k)}&=&\mathbf{A}_t^{(k)}\cdot(\mathbf{e}-\mathbf{A}^{(k)}_t)\cdot(\mathbf{e}-2\mathbf{A}^{(k)}_t)\nonumber\\
\mathbf{ddA}_{jt}^{(k)}&=&\bigg(\Big(\mathbf{W}^{(k+1)}\cdot \mathbf{A}_t^{(k)}\cdot(\mathbf{e}-\mathbf{A}^{(k)}_t)  \Big)\mathbf{ddA}_{jt}^{(k+1)}~'\bigg)'\label{ddA21}\\
&&+ \bigg(\Big\{\Big(\big(\mathbf{W}^{(k+1)}\cdot \mathbf{dda}_t^{(k)}\big)\mathbf{dA^b}_{t}^{(k+1)}\Big)\cdot \mathbf{W}^{(k)}~'\Big\}\mathbf{dA^f}_t^{k-1}~'\bigg)'\label{ddA22}
\end{eqnarray}
To see this, we note that \ref{da_b2} can be split into the product of $\mathbf{Z}_{1t}:=\mathbf{W}^{(k+1)}\cdot \mathbf{A}_t^{(k)}\cdot(\mathbf{e}-\mathbf{A}^{(k)}_t)$ and $\mathbf{Z}_{2t}:=\mathbf{dA^b}_t^{(k+1)}$. Therefore, its derivative can be split into the sum $\mathbf{dZ}_{1t}\mathbf{Z}_{2t}+\mathbf{Z}_{1t}\mathbf{dZ}_{2t}$: the first term corresponds to  \ref{ddA22} and the second term to \ref{ddA21}. In the former case $\big(\mathbf{W}^{(k+1)}\cdot \mathbf{dda}_t^{(k)}\big)\mathbf{dA^b}_{t}^{(k+1)}$ is the derivative of the outer function and the derivative of the inner function is accounted for by $ \mathbf{W}^{(k)}~'$ and $\mathbf{dA^f}_t^{k-1}~'$. Digging out dimensions, we first note that the matrix-product of the $(n_k,n_{k+1})$-dimensional $\mathbf{W}^{(k+1)}\cdot \mathbf{A}_t^{(k)}\cdot(\mathbf{e}-\mathbf{A}^{(k)}_t)$ and the $(n_{k+1},n)$-dimensional $\mathbf{ddA}_{jt}^{(k+1)}~'$ matrices in \ref{ddA21} has dimension $(n_k,n)$ so that the right-hand side of \ref{ddA21} has dimension $(n,n_k)$, after transposition. Similarly, the product of the $(n_k,n_{k+1})$-dim matrix $\mathbf{W}^{(k+1)}\cdot \mathbf{dda}_t^{(k)}$ and of the $n_{k+1}$-dim column-vector $\mathbf{dA^b}_{t}^{(k+1)}$ is a column-vector of dimension $n_k$ and hence $\Big(\big(\mathbf{W}^{(k+1)}\cdot \mathbf{dda}_t^{(k)}\big)\mathbf{dA^b}_{t}^{(k+1)}\Big)\cdot \mathbf{W}^{(k)}~'$ is of dimension $(n_k,n_{k-1})$ so that the product with the $(n_{k-1},n)$ dimensional $\mathbf{dA^f}_t^{k-1}~'$ has dimension $(n,n_k)$, after transposition, and we henceforth conclude that $\mathbf{ddA}_{jt}^{(k)}$ has dimension $(n,n_k)$. Note that this generic expression for $\mathbf{ddA}_{jt}^{(k)}$ simplifies at some particular layers, such as for example for $k=p-1$ (last hidden layer): in this case we consider the single output neuron $o_{jt}$ instead of the full output layer so that
\begin{eqnarray}
\mathbf{dda}_t^{(p-1)}&=&\mathbf{A}_t^{(p-1)}\cdot(\mathbf{e}-\mathbf{A}^{(p-1)}_t)\cdot(\mathbf{e}-2\mathbf{A}^{(p-1)}_t)\nonumber\\
\mathbf{ddA}_{jt}^{(p-1)}&=&\bigg(\Big(\mathbf{W}^{(p)}_j \cdot \mathbf{A}_t^{(p-1)}\cdot(\mathbf{e}-\mathbf{A}^{(p-1)}_t)\Big)\mathbf{ddA}_{jt}^{(p)}~' \bigg)' \label{ddA21i}\\
&&+ \bigg(\Big(\mathbf{W}^{(p)}_j {dA^b}_{jt}^{(p)}\cdot \mathbf{dda}_t^{(p-1)}\cdot \mathbf{W}^{(p-1)}~'\Big) \mathbf{dA^f}_t^{p-2}~'\bigg)'\label{ddA22i}
\end{eqnarray}
where ${dA^b}_{jt}^{(p)}$ is the $j$-th component of $\mathbf{dA^b}_{t}^{(p)}$, $\mathbf{W}^{(p)}_j$ is the $j$-th column vector of $\mathbf{W}^{(p)}$, of dimension $n_{p-1}$, and $\mathbf{ddA}_{jt}^{(p)}~'$ is a row-vector of dimension $n$ so that the right-hand side of \ref{ddA21i} is of dimension $(n,n_{p-1})$. On the other hand, the matrix $\mathbf{W}^{(p)}_j {dA^b}_{jt}^{(p)}\cdot \mathbf{dda}_t^{(p-1)}\cdot \mathbf{W}^{(p-1)}~'$ in \ref{ddA22i} inherits dimensions $(n_{p-1},n_{p-2})$ from $\mathbf{W}^{(p-1)}~'$ so that the entire expression in \ref{ddA22i} has dimension $(n,n_{p-1})$, after multiplication with $\mathbf{dA^f}_t^{p-2}~'$ and transposition. \\
In addition to the last hidden layer, the generic expression \ref{ddA22} simplifies also in the case of the first hidden layer $k=1$:
\begin{eqnarray}
\mathbf{dda}_t^{(1)}&=&\mathbf{A}_t^{(1)}\cdot(\mathbf{e}-\mathbf{A}^{(1)}_t)\cdot(\mathbf{e}-2\mathbf{A}^{(1)}_t)\nonumber\\
\mathbf{ddA}_{jt}^{(1)}&=&\bigg(\Big(\mathbf{W}^{(2)}\cdot \mathbf{A}_t^{(1)}\cdot(\mathbf{e}-\mathbf{A}^{(1)}_t)  \Big)\mathbf{ddA}_{jt}^{(2)}~'\bigg)'\label{ddA31}\\
&&+ \Big(\mathbf{W}^{(2)}\mathbf{dA^b}_{t}^{(2)}\cdot \mathbf{dda}_t^{(1)}\cdot \mathbf{W}^{(1)}~'
\Big)'\nonumber
\end{eqnarray}
whereby $\mathbf{dA^f}_t^{k-1}$ in \ref{ddA22} has been replaced by $\mathbf{dA^f}_t^{0}=\mathbf{Id}$ because the input-layer is not equipped with an activation function.  If the net is made of a single hidden layer then this is also the first as well as the last (hidden) layer so that the above simplifications can be merged. Finally, at the input layer $k=0$ we obtain
\begin{eqnarray}\label{ddda_b3}
\mathbf{ddA}_{jt}^{(0)}=\Big(\mathbf{W}^{(1)}\mathbf{ddA}_{jt}^{(1)}~'\Big)'=\mathbf{W}^{(1)}\mathbf{ddA}_{jt}^{(1)}~'=:\mathbf{QPD}_{jt}
\end{eqnarray}
where the $(n,n)$-dimensional $\mathbf{ddA}_{jt}^{(0)}$ is symmetric and corresponds to the QPD of the $j$-th output neuron at each time point $t$ i.e. $\mathbf{QPD}_{jt}$, $t=1,...,T$. As for the time-dependent $\mathbf{LPD}_t$, which differs from the traditional time-invariant parameter-gradient in backpropagation-algorithms, the time-dependent $\mathbf{QPD}_{jt}$ differs from the traditional time invariant parameter-Hessian found in optimization and inference, thus motivating the above derivations. \\

For the intercept defined by \ref{intercept} i.e. for
\begin{eqnarray*}
b_{jt}:=o_{jt}-\sum_{i=1}^n w_{ijt}x_{it}=o_{jt}-\mathbf{LPD}_{jt,-1}\mathbf{x}_{t}
\end{eqnarray*}
where $\mathbf{LPD}_{jt,-1}$ designates the ($t$-th row vector of) LPD without its first component, reserved for the intercept, and $\mathbf{x}_t$ is the $t$-th data column-vector, the corresponding 'QPD' can be derived from
\begin{eqnarray*}
\mathbf{\nabla b}_{jt}:=\mathbf{LPD}_{jt,-1}-\left(\mathbf{QPD}_{jt} \mathbf{x}_t+\mathbf{LPD}_{jt,-1}\right)=-\mathbf{QPD}_{jt} \mathbf{x}_t
\end{eqnarray*}
where $\mathbf{\nabla b}_{jt}$ is the gradient of the intercept with respect to the input data and where $\mathbf{LPD}_{jt,-1}$ and $\mathbf{QPD}_{jt}$ were derived previously.\\

Finally, as for the extension of the LPD in the previous section, second-order derivatives with respect  to arbitrary twice differentiable X-functions could be obtained  by substituting the composite activation function $xf(\sigma(\cdot))$ to $\sigma(\cdot)$ at the output neurons.  Specifically, differentiating the backward-expression \ref{da_b1xf} for the $j$-th output neuron we obtain the following extension of \ref{ddA}
\begin{eqnarray*}
\mathbf{dd(xf\circ A)}_{jt}^{(p)}&:=&{\nabla xf}({A}_{jt}^{(p)}) \mathbf{ddA}_{jt}^{(p)}+{\nabla^2 xf}({A}_{jt}^{(p)})\mathbf{dA^b}_{t}^{(p)} 
\end{eqnarray*}
where ${\nabla^2 xf}({A}_{jt}^{(p)})$ is the second-order derivative of $xf()$ at $o_{jt}={A}_{jt}^{(p)}$ and $\mathbf{dA^b}_{t}^{(p)}$ is defined in \ref{da_b1}. Proceeding backwards, through \ref{ddA21}, \ref{ddA31} and \ref{ddda_b3} then leads to the sought-after second-order sensitivities with respect to generic X-functions, as claimed.\\

To conclude note that the QPD is a measure of change  of the LPD as a function of the input data $\mathbf{x}_t$ at a fixed time point $t$ which is to be distinguished from changes of the LPD as a function of time $\mathbf{L}_{t}-\mathbf{L}_{t-1}$. In our examples below we will emphasize the latter, acknowledging that an analysis of the former might provide additional interesting insights.


\subsection{Alternative X-functions and Discrete Proxies}

A further potentially interesting X-function might be found in
\[
xf(o_t):=\frac{1}{T}\sum_{t=1}^T(y_t-o_t)^2=MSE
\]
where $y_t$ is the target series to be fitted by the net. The derived sensitivity
\[
dmse_{it}:=\partial xf(o_t)/\partial x_{it}
\]
can be interpreted as a measure for the importance of a data-point $x_{it}$, at time point $t$, in the determination of the proper net-parameters, i.e. weights and biases, as well as an assessment of potential overfitting of a particular data-point by the net or of a particular episode in the history of the data. The resulting data-flow IPD (Importance Parameter Data) allows to identify, to trace and to monitor time-points or -episodes of greater impact on the estimation criterion and thus on the internal structure of the net, as determined by its parameters. \\

Besides and in addition to the above vanilla X-functions, we may also consider context-specific or customized X-functions, such as generic performance measures: in the invoked financial context the X-function could be identified with the Sharpe ratio or with a risk-measure (loss-quantile,  maximal drawdown) or with any measure summarizing the rationale of a particular investor's perspective. Besides a proper interpretability of the net, in terms of marginal contribution of a data-point to the chosen performance measure, the resulting time series of sensitivities could be used to identify data-points or time episodes of higher impact on the aggregate performance.
Note, however, that arbitrary general  X-functions might not be differentiable anymore. 
In such a case, the discrete proxy
\[
\Delta_{it}^{xf}(\delta|\delta>0):=\displaystyle{\frac{xf(x_{it}+\delta)-xf(x_{it})}{\delta}}
\]
might provide a valuable alternative. Also, the proposed extension could be applied to discrete-valued data, for example classes, or eventually to alternative machine learning techniques. However, discrete changes introduce new artificial data $x_{it}+\delta$ which might potentially conflict with the local dependence structure of the data as reflected by the conditional distributions $f(x_{it}|x_{kt},k\neq i)$; moreover, discrete derivatives $\Delta_{it}^{xf}(\delta|\delta>0)$ are reliant on the selection of $\delta$ as well as on numerical precision (numerical cancellation); finally, higher order derivatives such as the above QPD could heavily magnify these issues.\\

\subsection{Analyzing and Monitoring the Entire Net-Structure in Real-Time}

In the previous sections our intent was to derive 'explainability' by analyzing sensitivities with respect to \emph{input} data. However, the above computations of LPD or QPD could be interrupted at any hidden-layer $k$, $1\leq k\leq p-1$: collecting these intermediate sensitivities can inform about the importance (weight) or the non-linearity or the impact on overfitting or the contribution to the trading performance of each neuron at each time point $t$. 
In addition to the proper explainability aspect, this data points at the state of the net in real-time and, by extension, at possible 'states' of the data generating process. We now proceed by illustrating the new XAI-tool, as based on the LPD.

\section{Applications of the LPD}

\begin{figure}[H]\begin{center}\includegraphics[height=3in, width=3in]{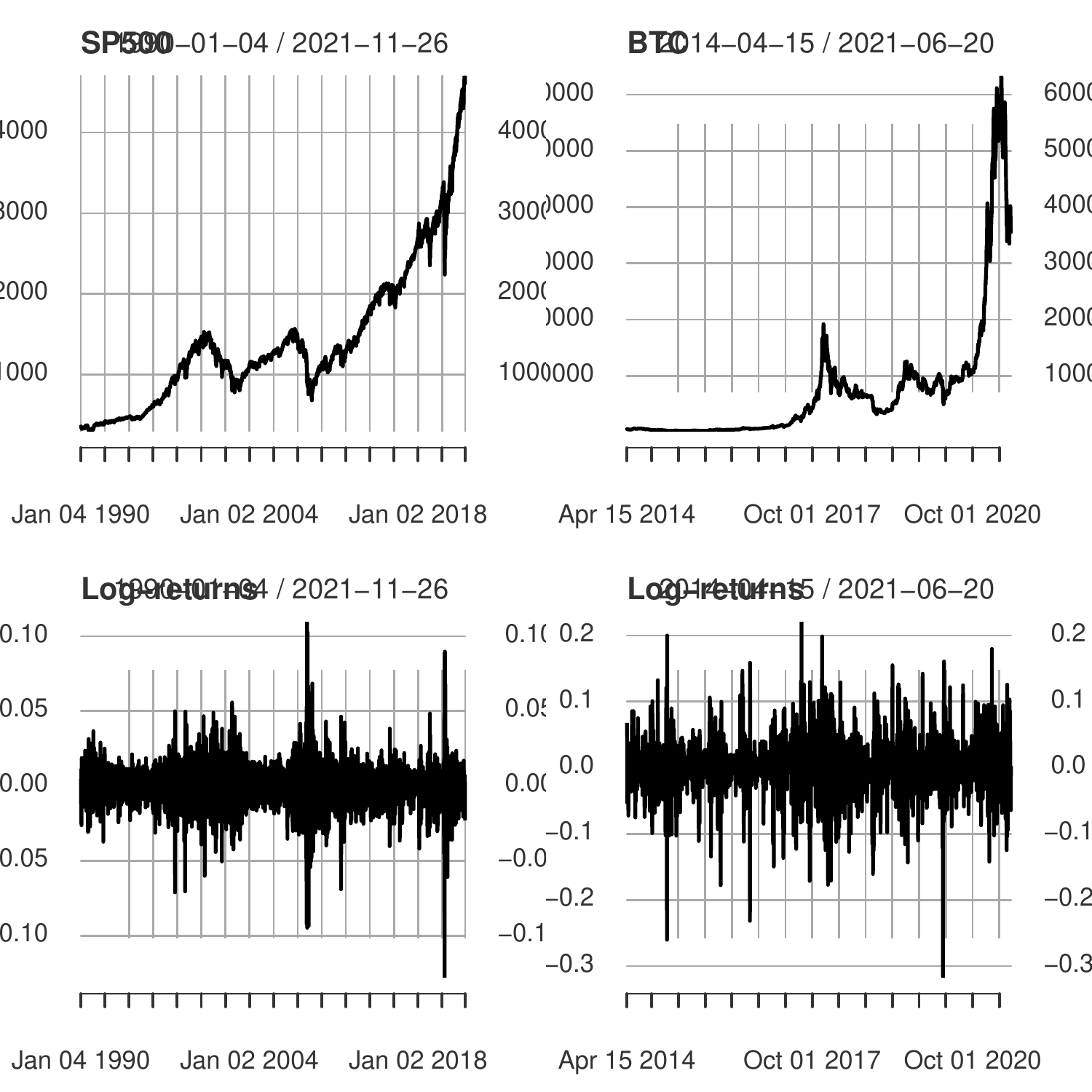}\caption{SP500 (left panels) and BTC (right panels): prices (upper panels) and log-returns (bottom panels)\label{data}}\end{center}\end{figure}

In the sequel we apply the LPD to the Bitcoin (BTC) crypto-currency and to the S$\&$P 500 equity-index: both series are displayed in fig.\ref{data}.
The price series (top panels) are non-stationary and the log-returns (bottom panels) show evidence of vola-clustering or conditional heteroscedasticity. Furthermore, the BTC is subject to more frequent extreme events and more pronounced up- and down-turns. 
The LPD in these examples will be based on a simple neural net architecture, introduced in \cite{bundiwildi20},  applied to the log-returns of the series (an application to price-data will be considered too, for the equity-index).  
Besides and in addition to the proper explainability aspect, an application of simple statistical analysis to the resulting LPD series will suggest possible extensions of the framework to Risk-Management (RM) and Fraud-Detection (FD). Note that data is always mapped to the unit-interval when fitting net-parameters but results, such as forecasts or trading-performances, are always transformed-back to original prices or log-returns.

\subsection{BTC}

\begin{figure}[H]\begin{center}\includegraphics[height=3in, width=3in]{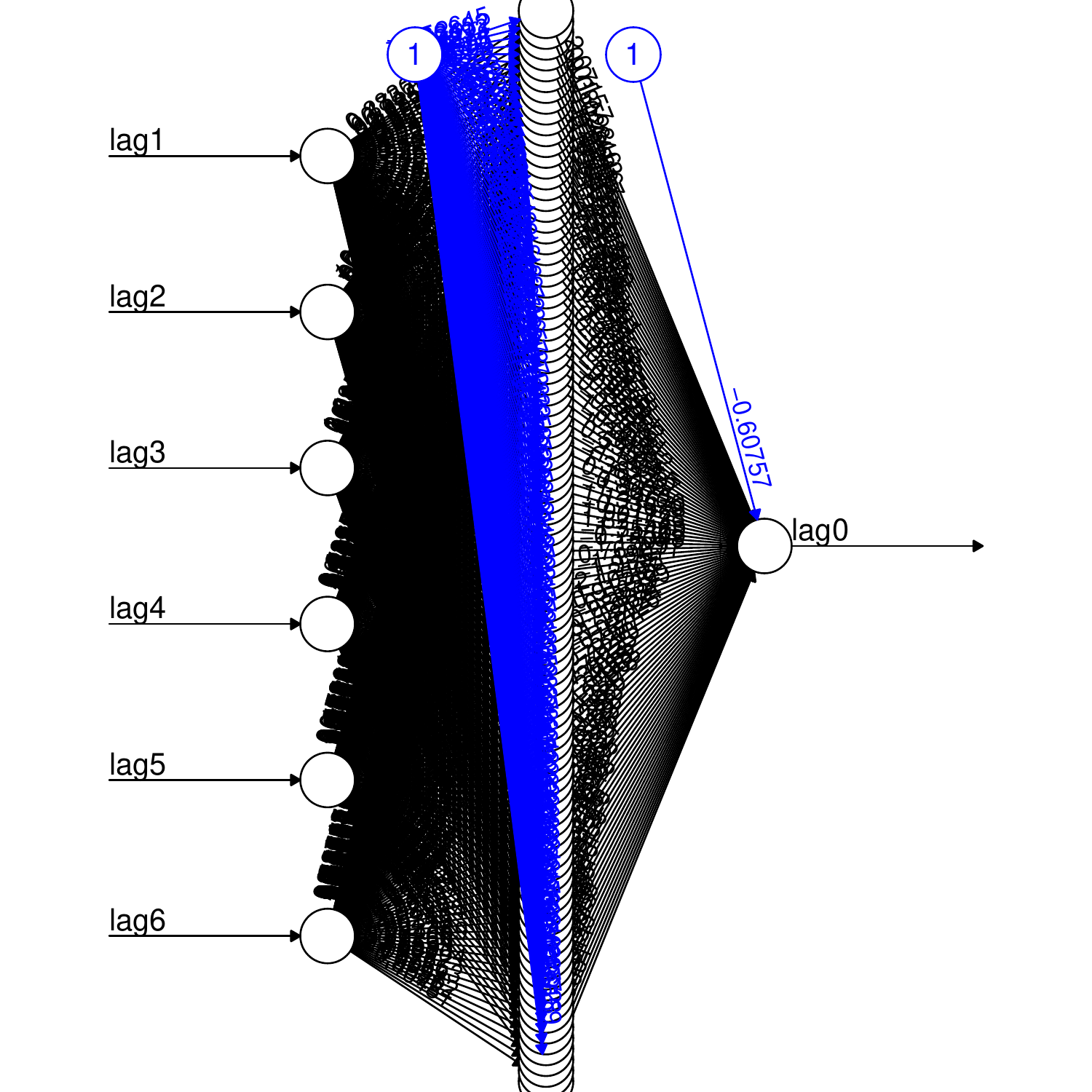}\caption{Neural net BTC: feedforward net with a single hidden-layer of dimension 100 and an input layer of dimension 6 comprising the last six lagged returns\label{nn_100}}\end{center}\end{figure}

We rely on \cite{bundiwildi20} where the authors propose a simple feedforward net with a single hidden layer and an input layer collecting the last six lagged (daily) returns: the net is then trained to predict next day's return based on the MSE-criterion. We here slightly deviate from the proposed specification by proposing a richer parameterized net with a hidden layer of dimension one hundred in order to highlight explainability aspects. The number of estimated parameters then amounts to a total  of $6*100+100=700$ weights and $100+1=101$ biases, see fig.\ref{nn_100}.
The in-sample span for estimation or 'learning' covers an episode of roughly four years, from the first quotation on the Bitstamp crypto-exchange, in 2014, up to a peak of the currency in December 2017. The selected out-of-sample span is then subject to severe draw-downs and strong recoveries which provide ample opportunity for 'smart' market-positioning. \\

To begin our analysis of the BTC, we now propose a simple solution for handling the numerical indeterminacy (random-net) illustrated in section \ref{r_toy}.

\subsubsection{Random Nets and Random Trading Performances }

\begin{figure}[H]\begin{center}\includegraphics[height=3in, width=3in]{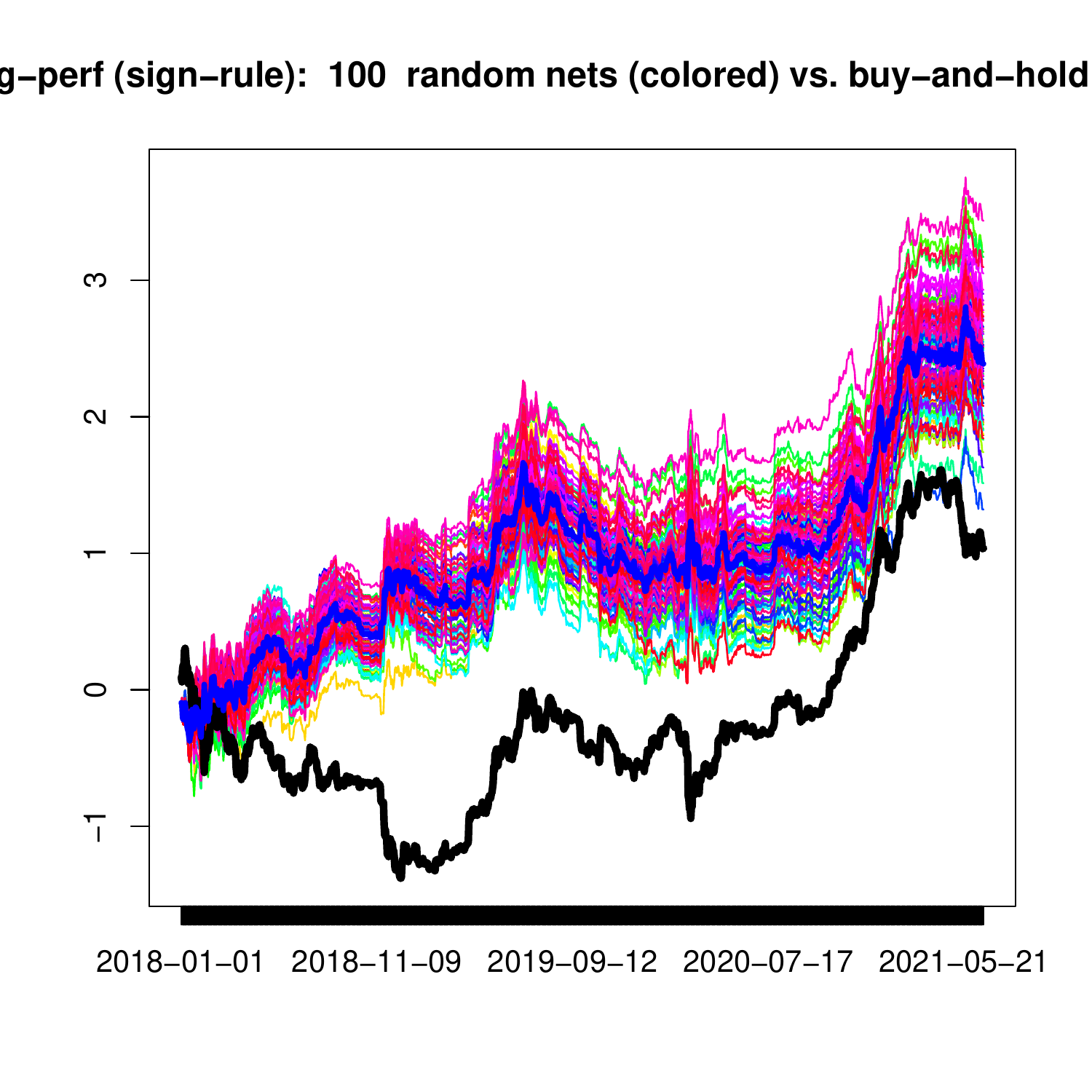}\caption{Cumulated log-performances out-of-sample based on sign-rule (buy or sell depending on sign of forecasted return): 'random' neural nets (colored) vs. buy-and-hold (bold black) and mean-net performance (bold blue) \label{bit_perf_sign_random}}\end{center}\end{figure}

We here optimize the net 100-times, based on different random initializations of its parameters, and we compute trading performances of each random-net based on the simple sign-rule: buy or sell tomorrow's return depending on the sign of today's net-forecast. The resulting cohort of cumulated (out-of-sample) trading performances is displayed in fig.\ref{bit_perf_sign_random} with the mean-performance in the center (bold blue line). Remarkably, even the least performing net outperforms the buy-and-hold benchmark (lower black line) in the considered out-of-sample span and the net-cohort systematically mitigates draw-downs of the BTC at the expense of slightly weaker  growth during hefty upswings: this issue can be addressed later, when exploiting the LPD.
While fig.\ref{bit_perf_sign_random} suggests a fairly broad range of 'random' trading-realizations, the aggregate mean has stabilized and is virtually invariant to the particular random-seed selected for parameter initialization. 
We now address the problem of picking-out the best possible out-of-sample random-net based on historical in-sample performances. For this purpose table \ref{in_out_perf} reports correlations between in-sample and out-of-sample forecast and trading-performances: a strong correlation indicates that the best in-sample net is likely to perform well out-of-sample, too. The table suggests that out-of-sample trading performances are not directly related to in-sample forecast- or trading-performances so that the aggregate mean, assigning equal weight to each of the 100 random-nets in fig.\ref{bit_perf_sign_random} is a valuable strategy, at least in the absence of further empirical evidences; moreover this strategy is virtually invariant to the random seed of parameter initialization(s).
\begin{table}[ht]
\centering
\begin{tabular}{rrrr}
  \hline
 & mse in/mse out & mse in/Sharpe out & Sharpe in/Sharpe out \\ 
  \hline
Correlation & 0.894 & -0.115 & -0.138 \\ 
   \hline
\end{tabular}
\caption{Correlations between in-sample and out-of-sample performances} 
\label{in_out_perf}
\end{table}
We now proceed by computing random-LPD and aggregate mean-LPD,  aiming for a better understanding of the richly parameterized net(s).

\subsubsection{Interpretability: LPD, Variable-Relevance and an Unassuming Forecast Heuristic}

\begin{figure}[H]\begin{center}\includegraphics[height=3in, width=3in]{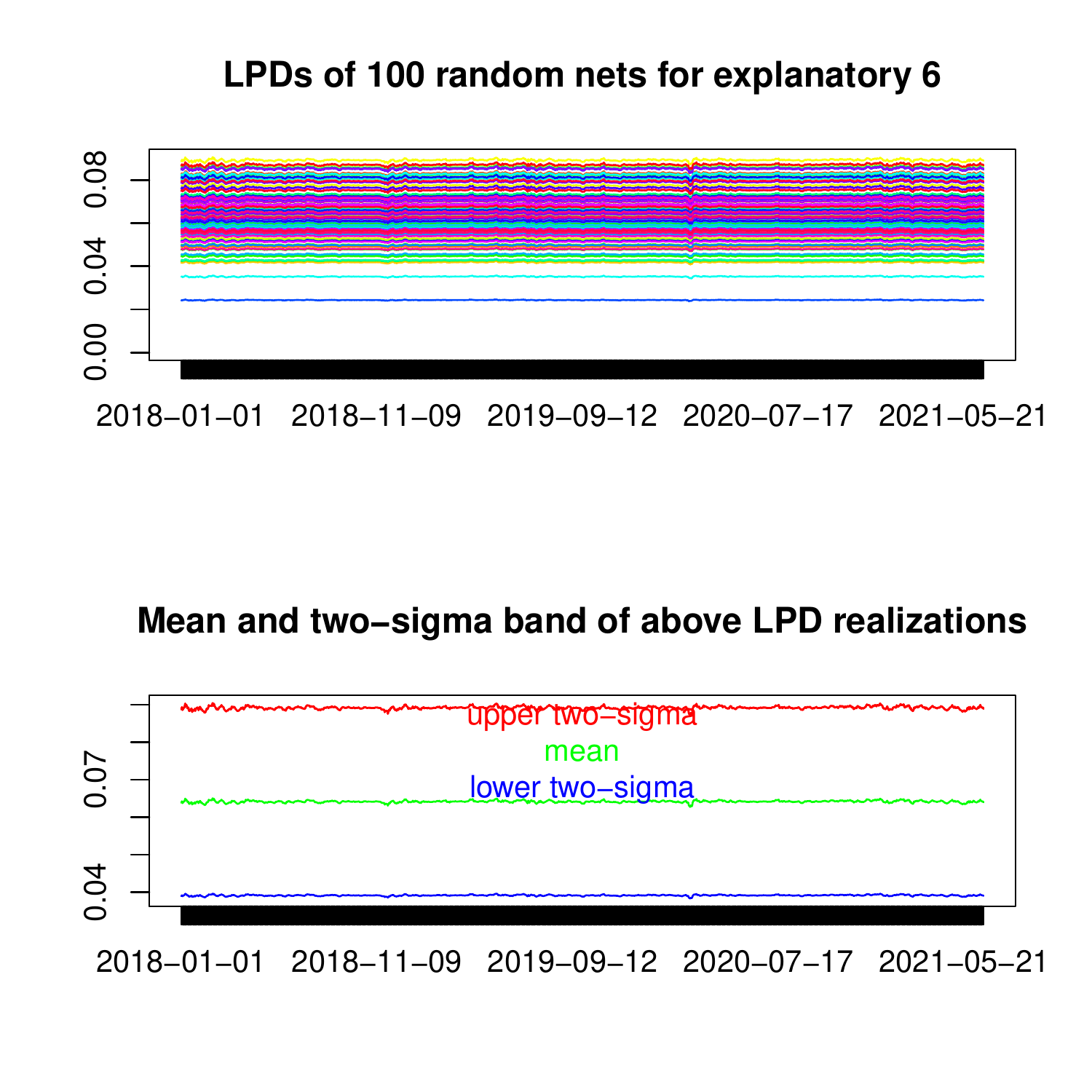}\caption{Out-of-sample random LPDs of the lag-6 BTC input variable (top) and mean-LPD with  empirical two-sigma band (bottom)\label{LPD_array_out_sample_2sigma}}\end{center}\end{figure}

\begin{figure}[H]\begin{center}\includegraphics[height=3in, width=3in]{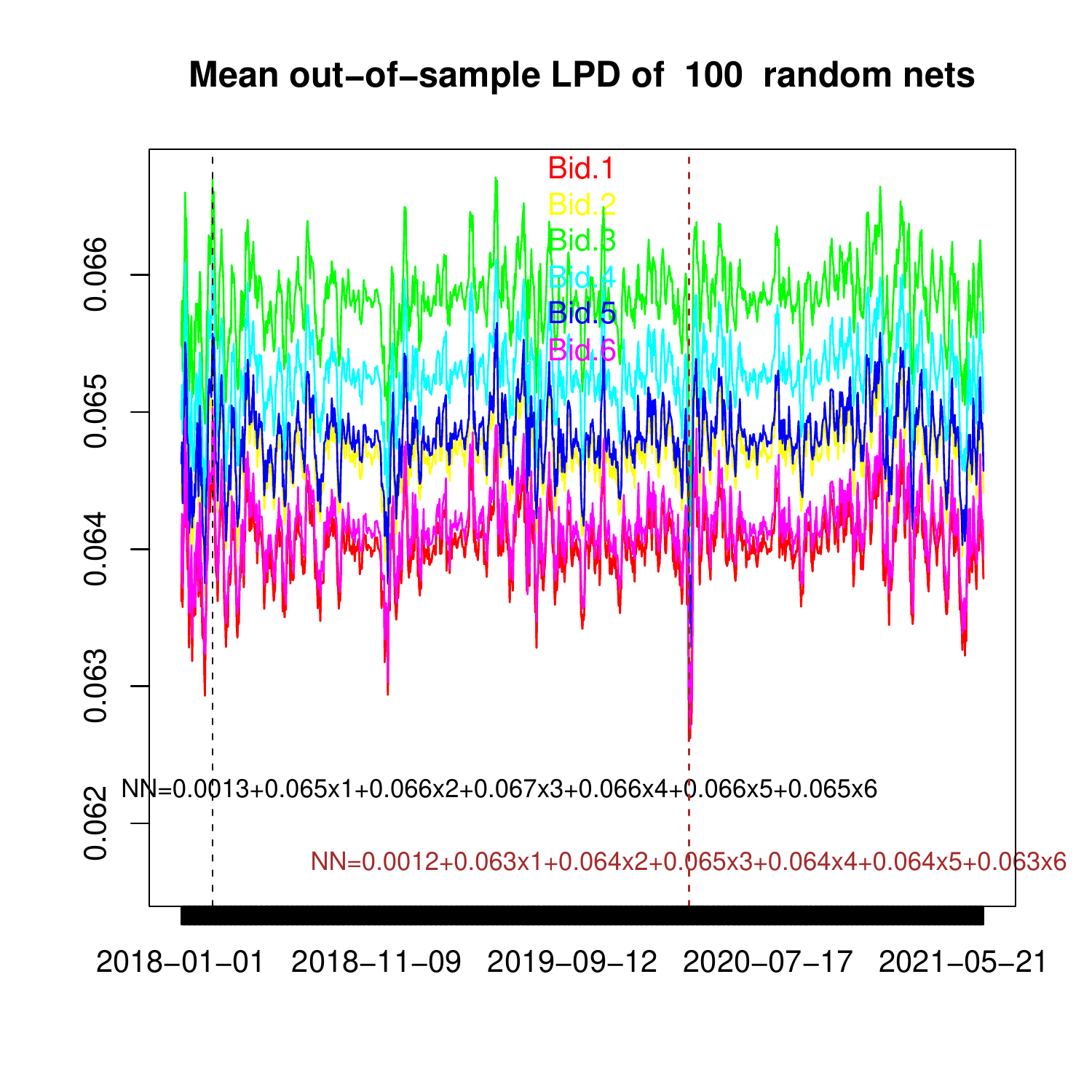}\caption{Out-of-sample mean LPD: means are taken for each explanatory variable over all random nets\label{LPD_array_out_sample_agg}}\end{center}\end{figure}

\begin{figure}[H]\begin{center}\includegraphics[height=3in, width=3in]{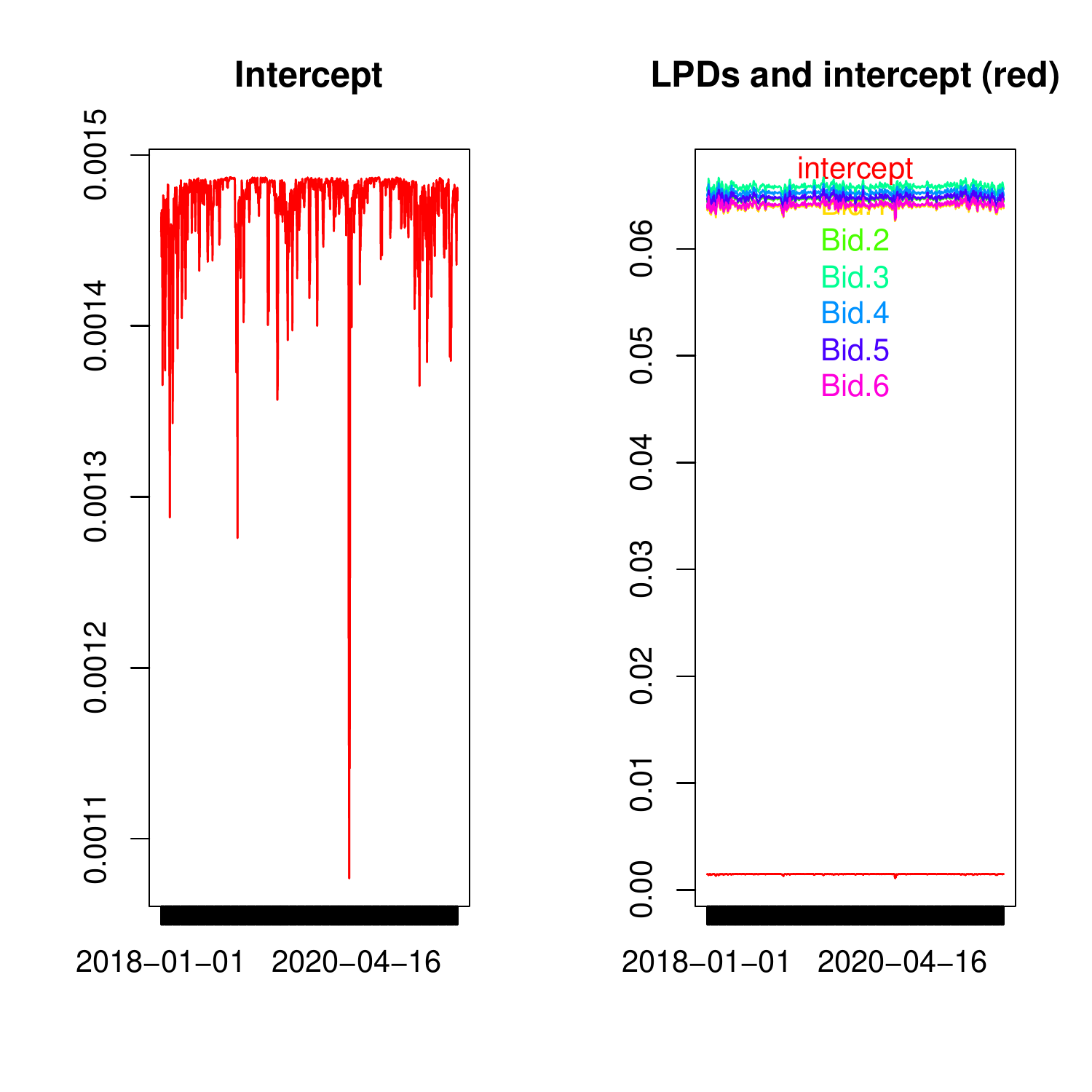}\caption{Out-of-sample mean intercept (left panel) and all LPDs plus intercept (right panel)\label{LPD_array_out_sample_agg_inter}}\end{center}\end{figure}

For ease of exposition we first display the random realizations of the last column of the LPD only, corresponding to the lag-six return $r_{t-6}$, see fig.\ref{LPD_array_out_sample_2sigma}, upper panel\footnote{The lag-six return is identified as an important explanatory variable in \cite{bundiwildi20}.}. The lower panel in the figure displays the corresponding mean-LPD $\overline{LPD}_{t,6}$, together with empirical two-sigma bands $\overline{LPD}_{t,6}\pm 2\sigma_{t,6}$, where
\begin{eqnarray}
\overline{LPD}_{t,6}&:=&\frac{1}{100}\sum_{i=1}^{100} LPD_{t,6i}\nonumber\\
\sigma_{t,6}&:=&\frac{1}{100}\sum_{i=1}^{100} (LPD_{t,6i}-\overline{LPD}_{t,6})^2\label{sigma_lpd}
\end{eqnarray}
and  where $LPD_{t,6i}$ is the LPD of the lag-6 variable of the $i$-th random net. Since the LPD corresponds to the parameters of a (time-dependent) linear replication of the net, synthetic t-statistics could be computed for inferring the relevance of the  explanatory variables by computing the ratio of mean-LPD and standard-deviation
\[
\mathbf{t}_t:=\displaystyle{\frac{\overline{\mathbf{LPD}}_t}{\boldsymbol{\sigma}_t}}
\]
at each time point $t$, corresponding to (a vector of) synthetic t-statistics, one for each input variable\footnote{The proposed relevance concept differs from classic statistical significance because the random process affects initial values of parameters, which is to be distinguished from the classic data generating process (formal details are omitted here). }.
Interestingly, given the richly parameterized net structure, the plot suggests nearly constant sensitivities along the time axis, as could be ascribed to an ordinary linear model, at least up to the random-effects due to parameter initialization;  
moreover, these findings are confirmed across all explanatory variables, as can be seen in fig.\ref{LPD_array_out_sample_agg} which displays the corresponding mean LPDs; in addition, the last figure reveals that the variables receive nearly equal weight, in the mean. To complete the analysis, fig.\ref{LPD_array_out_sample_agg_inter} displays the mean intercept defined by \ref{intercept}, either alone (left panel) or together with the above mean LPDs (right panel). In order to summarize our findings we now  introduce the mean net-output
\[
\overline{o}_t=\frac{1}{100}\sum_{i=1}^{100} o_{t,i}=\overline{\mathbf{LPD}}_t \left(\begin{array}{cc}&1\\&\mathbf{x}_t\end{array}\right)\approx 0.0015 +0.065\sum_{j=1}^6 x_{jt}
\]
where $\overline{\mathbf{LPD}}_t$ is the vector of mean-intercept and mean-LPDs of fig.\ref{LPD_array_out_sample_agg_inter}.
We then infer that the consensus-forecast $\overline{o}_t$ of the random-nets can be approximated by an unassuming equally-weighted MA(6) forecast-heuristic, with constant weights roughly equal to $0.065$, shifted by a small intercept of size $0.0015$. In this sense, the proposed mean-LPD effectively resolves both the black-box non-linearity of the richly parameterized net as well as the indeterminacy and randomness of  parameter estimates. Interestingly, the equally-weighted MA(6) was already identified as a successful strategy for the BTC in  \cite{bundiwildi20}. As a result of the above explainability effort, we can ascribe trustworthiness to the consensus forecast of the armada of random-nets by relating it to a simple forecast heuristic; regrettably, the added-value in terms of pure forecast-gains  is rather underwhelming. Nonetheless, the nets can be exploited differently, as proposed in the following sections.


\subsubsection{First-Order Linear Approximation and Second Order Common Non-Linerarity}

In the previous section we emphasized a linear first-order approximation of the net for the purpose of explainability. In addition and in complement, we here briefly analyze second-order non-linearity, as measured by deviations of the LPD from constant. Fig. \ref{LPD_array_out_sample_agg} suggests that second-order non-linearity takes the form of a stationary process, strongly correlated across explanatory variables, as confirmed by table \ref{cor_LPD}, first row; moreover, 'random' non-linearity, overlaying random-LPDs, is also  strongly cross-correlated with the mean and therefore across variables, see the second row of the table. 
\begin{table}[ht]
\centering
\begin{tabular}{rrrrrrr}
  \hline
 & Lag 1 & Lag 2 & Lag 3 & Lag 4 & Lag 5 & Lag 6 \\ 
  \hline
Correlation across mean LPDs & 1.000 & 0.981 & 0.983 & 0.983 & 0.983 & 0.982 \\ 
  Correlation of mean and random LPDs & 0.986 & 0.991 & 0.992 & 0.992 & 0.993 & 0.992 \\ 
   \hline
\end{tabular}
\caption{Cross-correlations of mean LPDs (first row: correlations referenced against the lag-1 variable) and of random LPDs (second row: correlations  referenced against the corresponding means). For each explanatory variable the mean of the 100 random-correlations are reported in the second row.} 
\label{cor_LPD}
\end{table}In contrast to the first-order idiosyncratic random effects, inherited by parameter initialization, second-order non-linearity appears as a common factor, pervading the LPD in all dimensions. We here conjecture that this 'non-linearity' factor originates in the data or, more precisely, in the data-generating process, which is common to all random-nets, see further evidences in the following empirical sections. 
By hinting at changes in the data-generating process, the common (second-order non-linearity) factor can complement the first order linear approximation, chiefly emphasized in the previous section, by gathering a better understanding  of the data from a more refined analysis of the model, as explained in the next sections.

\subsubsection{Risk-Management: Exploiting Second Order Information of the LPD}

\begin{figure}[H]\begin{center}\includegraphics[height=3in, width=3in]{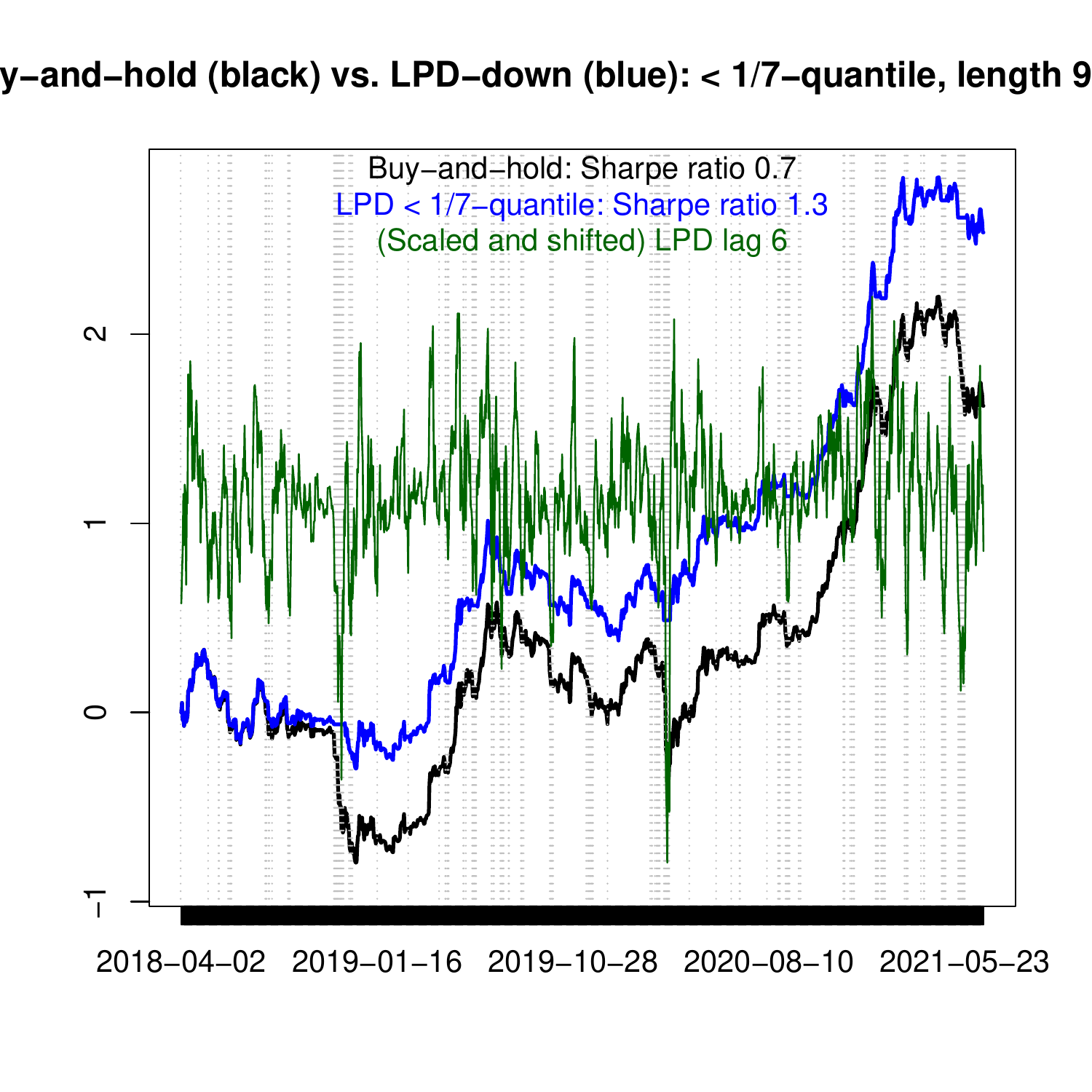}\caption{Buy-and-hold (black) vs. out-of-sample (mean-) LPD market-exit strategy (blue): exits (shaded in grey) occur if today's out-of-sample (absolute) mean-LPD (green) drops below the 1/7-quantile based on a rolling-window of length one quarter of its own history. The LPD corresponding to the lag-6 BTC-value is used.\label{LPD_adjusted_lower_90}}\end{center}\end{figure}

\begin{figure}[H]\begin{center}\includegraphics[height=3in, width=3in]{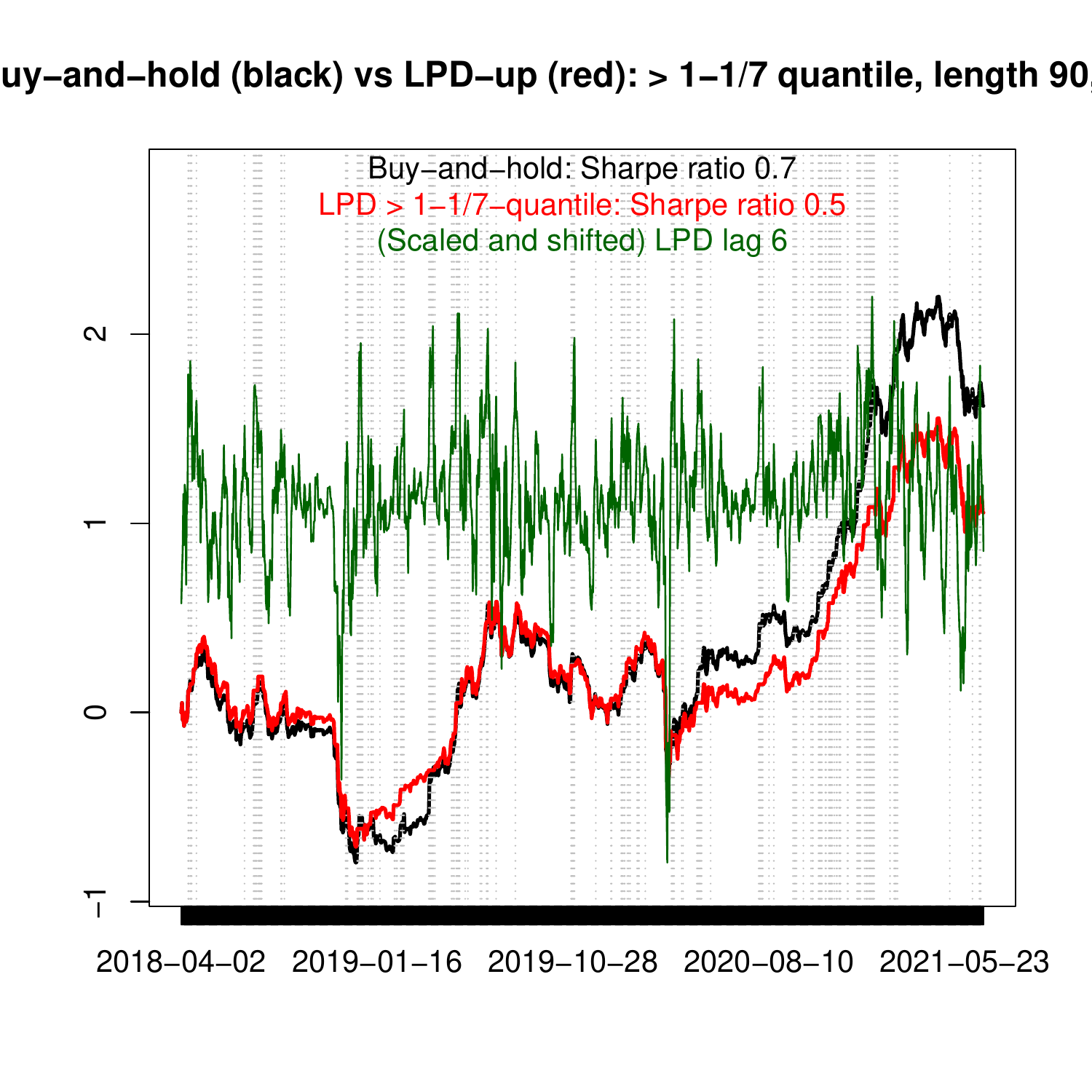}\caption{Buy-and-hold (black) vs. Out-of-sample (mean-) LPD market-exit strategy (red): exits (shaded in grey) occur if today's out-of-sample (absolute) mean-LPD (green) exceeds the upper 1-1/7 quantile based on a rolling-window of length one quarter of its own history. The LPD corresponding to the lag-6 BTC-value is used.\label{LPD_adjusted_upper_90}}\end{center}\end{figure}

\begin{figure}[H]\begin{center}\includegraphics[height=3in, width=3in]{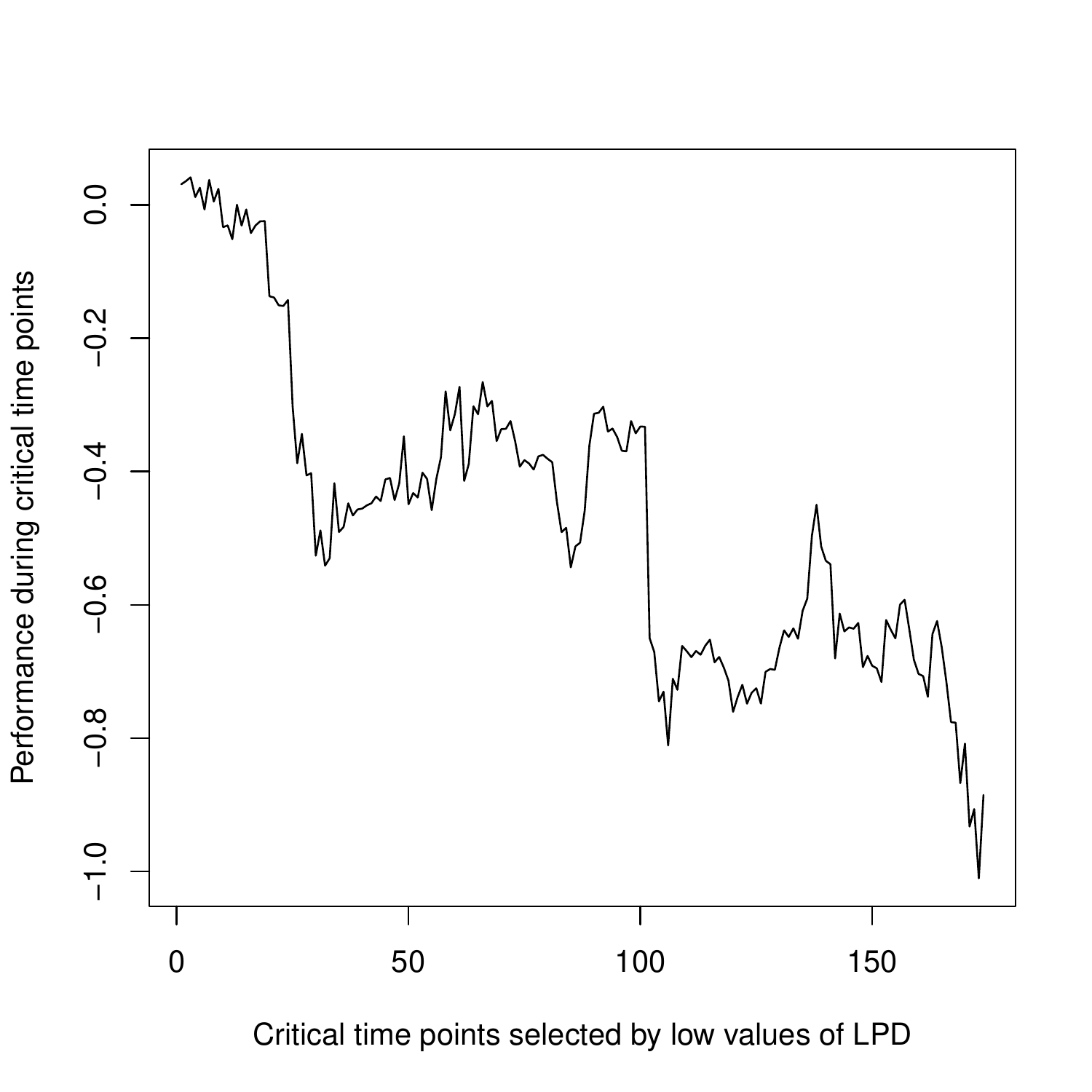}\caption{Drift of next day's BTC at critical time points identified when the (absolute) LPD drops below the 1/7-quantily: a negative drift confirms correct identification of BTC-downturns by the strategy\label{Drift_LPD_adjusted_month}}\end{center}\end{figure}

While it is often understood that risk-management (RM) concerns a mitigation of risk by means of diversification (central limit theorem),  
we here propose a more direct approach towards risk by identifying 'uncertain times' or 'unusual states' of the market in real-time.
\begin{itemize}
\item Unusual states are identified by episodes during which the dependence structure, as measured by the LPD, is weak.
\item Alternatively, unusual states are identified by episodes of increased non-linearity i.e. by large QPD realizations.
\item Uncertain times are identified by episodes during which the random nets are inconclusive about the local linear model i.e. times at which the standard deviations  $\boldsymbol{\sigma}_t$ of LPD (or QPD), defined in \ref{sigma_lpd}, are large.
\end{itemize}
Due to space limitations, we here emphasize the first strategy only 
and in order to simplify exposition we further restrict attention to the lag-six LPD, acknowledging that the relevant second order information is heavily redundant across variables, see table \ref{cor_LPD}. In this context, a market-exit (cash-position) is triggered when the dependence structure of the data, as measured by the LPD, is weakening or, in other words, when the net-forecast is less conclusive about next-day's return. In order to formalize the 'weakness' concept we assume first that exit-signals have a probability of 1/7, one per week in the mean, which reflects the frequent occurrence of local bursts or disruptions of the BTC: a market exit is then triggered if the \emph{absolute value} of the LPD drops below its empirical $1/7$-quantile. The computation of the quantile is based on a rolling-window of length 100 days, corresponding roughly to the last quarter of observations: we argue that a quarter of data is sufficiently long for resolving the corresponding tail of the distribution, at least with respect to the 1/7-quantile, and it is short enough to adapt for possible structural changes in the BTC\footnote{The selection of these parameters is not critical since alternative settings, further down, will result in qualitatively comparable outcomes.}. 
Finally, we cross-check the proposed RM monitoring-tool by analyzing crossings of the mirrored 1-1/7 upper quantile by the LPD, corresponding to unusually strong dependence, see 
figures \ref{LPD_adjusted_lower_90} and \ref{LPD_adjusted_upper_90}. A direct comparison of market-exits in the figures suggests 
that \emph{weak data dependence} (LPD below lower quantile) matches by the majority \emph{down-turns} of the BTC and conversely \emph{strong data dependence} (LPD above upper quantile) matches by the majority \emph{up-turns}, thus confirming the original intent and rationale of the proposed RM-strategy. Further empirical evidences are proposed for alternative quantile specifications, see fig.\ref{LPD_two_sided_signals}, and a different asset, the S$\&$P500-index, see section \ref{sp500sec}. We may argue that (asymmetric) markets dominated by long-positioned investors become less systematic during loss-phases due to possible herding or panic, thus establishing a link between (weak) dependence and (weak future) growth, see table \ref{crit_LPD} and the corresponding analysis below.
 
The suggested concomitance of negative BTC-growth and of weak data-dependence is illustrated  further in  fig.\ref{Drift_LPD_adjusted_month} which displays cumulated next-day's returns during critical time points tagged as exit-signals in fig.\ref{LPD_adjusted_lower_90}: the negative drift is systematic and strong, against and despite of 
the positive long-term drift of BTC. Note also that a corresponding long-short strategy, being  short-positioned at critical time points and long-positioned otherwise, would outperform substantially the risk-averse strategy in fig.\ref{LPD_adjusted_lower_90}, where market-exits cause a flattening of the performance line which ought to be desirable depending on the risk-profile. This long-short strategy would also outperform by a fair margin the previous neural net benchmarks, in fig.\ref{bit_perf_sign_random},  suggesting that second-order non-linearity, as formalized by changes of the LPD, can complement simple forecast-based trading-rules.
\begin{table}[ht]
\centering
\begin{tabular}{rll}
  \hline
 & Proportion of positive signs & Average next days' returns \\ 
  \hline
Critical time points & 51.1\% & -0.509\% \\ 
  Neutral time points & 53.8\% & 0.244\% \\ 
  Auspicious time points & 45.9\% & 0.305\% \\ 
  All time points & 52.2\% & 0.142\% \\ 
   \hline
\end{tabular}
\caption{Proportions of positive signs and average next days' returns based on critical time points (|LPD| < lower quantile: weak dependence), neutral time points (lower quantile < |LPD| < upper quantile: normal dependence), auspicious time points (|LPD| > upper quantile: strong dependence) and all time points.  } 
\label{crit_LPD}
\end{table}Table \ref{crit_LPD} provides additional evidence and alternative insight about the connection between the LPD and next day's return: the LPD is not conclusive about the sign of next-day's return (first column) but about the sign of the \emph{average} next day return (second column) i.e. the LPD supports information about the skewness of the distribution. As an example, the table suggests that 
the average return during critical time points, namely -0.509 (first row, second column), exceeds by a multiple of 3.6, in absolute value, the already impressive mean-drift 0.142  of the BTC over the entire time span (fourth row, second column). The pronounced skewness might point at the intermittent occurrence of panic-selling at time points tagged by low LPD-values. Indeed, the former  phenomenon is likely to 'cause' intermittently lower dependence structure of the data, thus establishing a link between LPD and average return which could serve as an 'explanation' or at least a justification for the proposed RM approach.  As a further evidence, episodes of strong dependence, as tagged by large (absolute) LPD values, support a positive average return of 0.305, largely in excess of the mean BTC-return over the entire time span. \\

\subsubsection{Fraud Detection: Exploiting Second Order Information of the LPD}

\begin{figure}[H]\begin{center}\includegraphics[height=3in, width=3in]{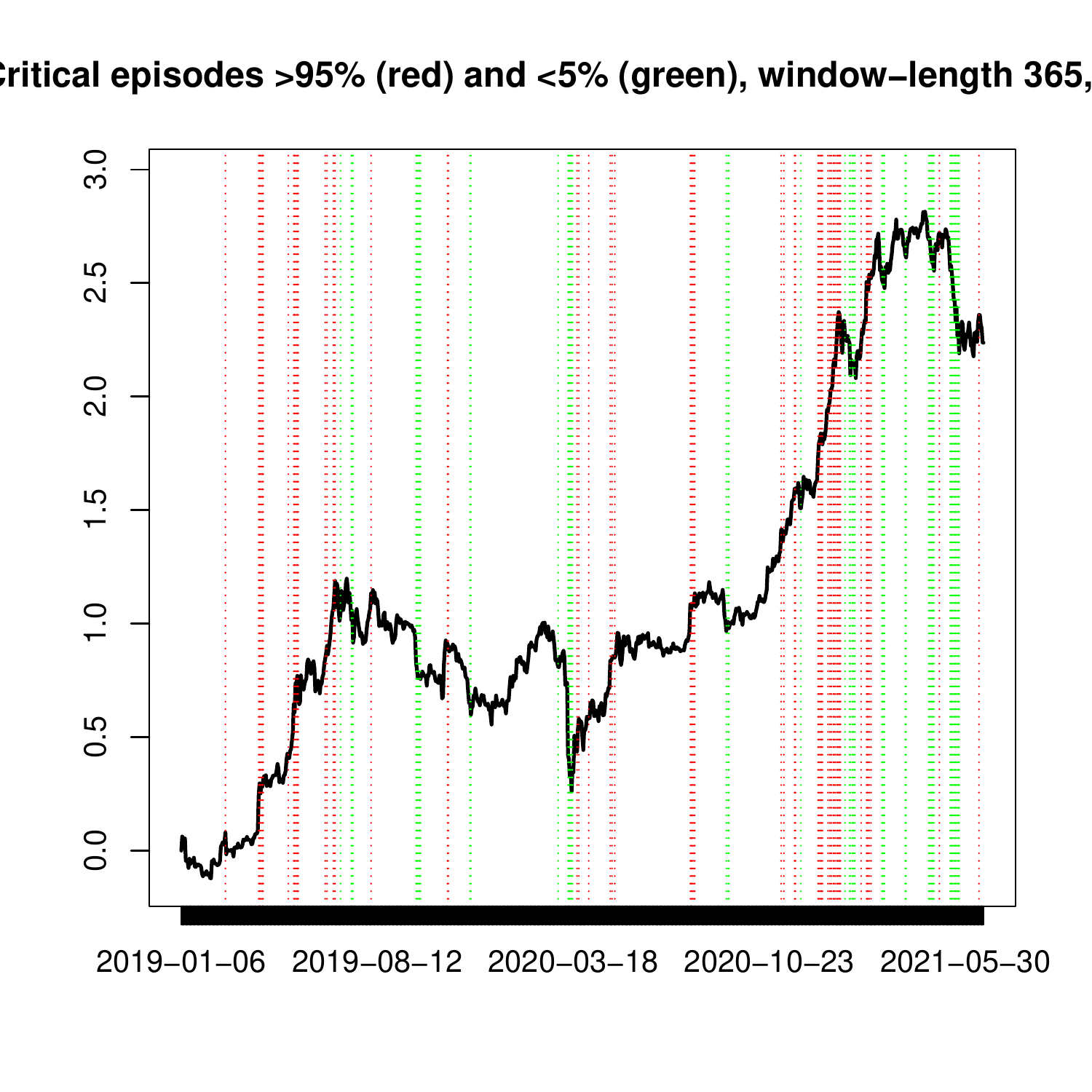}\caption{Critical time points: two-sided exceedances of the 5-percent (green) and 95-percent (red) quantiles of the  out-of-sample mean-LPD (mean over 100 nets) based on a rolling-window of length one year of its own history. The LPD corresponding to the lag-6 BTC-value is used.\label{LPD_two_sided_signals}}\end{center}\end{figure}

We here try to address situations or episodes of 
strongly unusual activity in the BTC-market, which could be more typically related to the field of fraud detection. 
In contrast to the previous RM-framework, we then propose a more stringent rule by selecting the lower $5\%$ quantile; moreover, we consider two-sided exceedances i.e. either unusually strong (|LPD|>1-1/20 quantile) or weak (|LPD|<1/20 quantile) dependence structures; finally, the length of the rolling window for computing the quantiles is increased to a full year in order to obtain sufficient resolution of the tail distribution. Fig.\ref{LPD_two_sided_signals} displays lower (green) and upper (red) quantile exceedances: 
once again, weak dependence seems mostly related to down-turns of the BTC and conversely for strong dependence.
We here suggest that these episodes are possibly indicative of unusual activity that might call for accrued attention and care of investors or regulators: as an example the last two green and red triggers, to the right-most of the plot, correspond to well-documented singular social media events (twitter messages): the first (May 2021) in support  and the second against the BTC (June 2021). In this context, we argue that the LPD could provide additional and alternative insights about the increasingly relevant application-field referred to as 'fraud detection'. While classic approaches to fraud detection often rely on residual analysis, tagging time points at which a model's output deviates excessively from target, the LPD does not rely on a target at all; rather it is the state of the net, as summarized by the LPD, which is indicative of 'suspect' data in terms of excessively small or large sensitivities of the net with respect to a X-function.\\ 


\subsection{S$\&$P 500}\label{sp500sec}

We here cross-check some of our earlier findings by relying on the S$\&$P 500 index and we take opportunity to discuss the effects of non-stationary trends by fitting nets to log-returns as well as to price data: the latter is mapped to the unit-interval for parameter-fitting but all results are transformed back to original scales and levels. Finally, we also consider the synthetic intercept \ref{intercept} as a means for identifying market-risks in real-time.

\subsubsection{Random Performances}

\begin{figure}[H]\begin{center}\includegraphics[height=3in, width=3in]{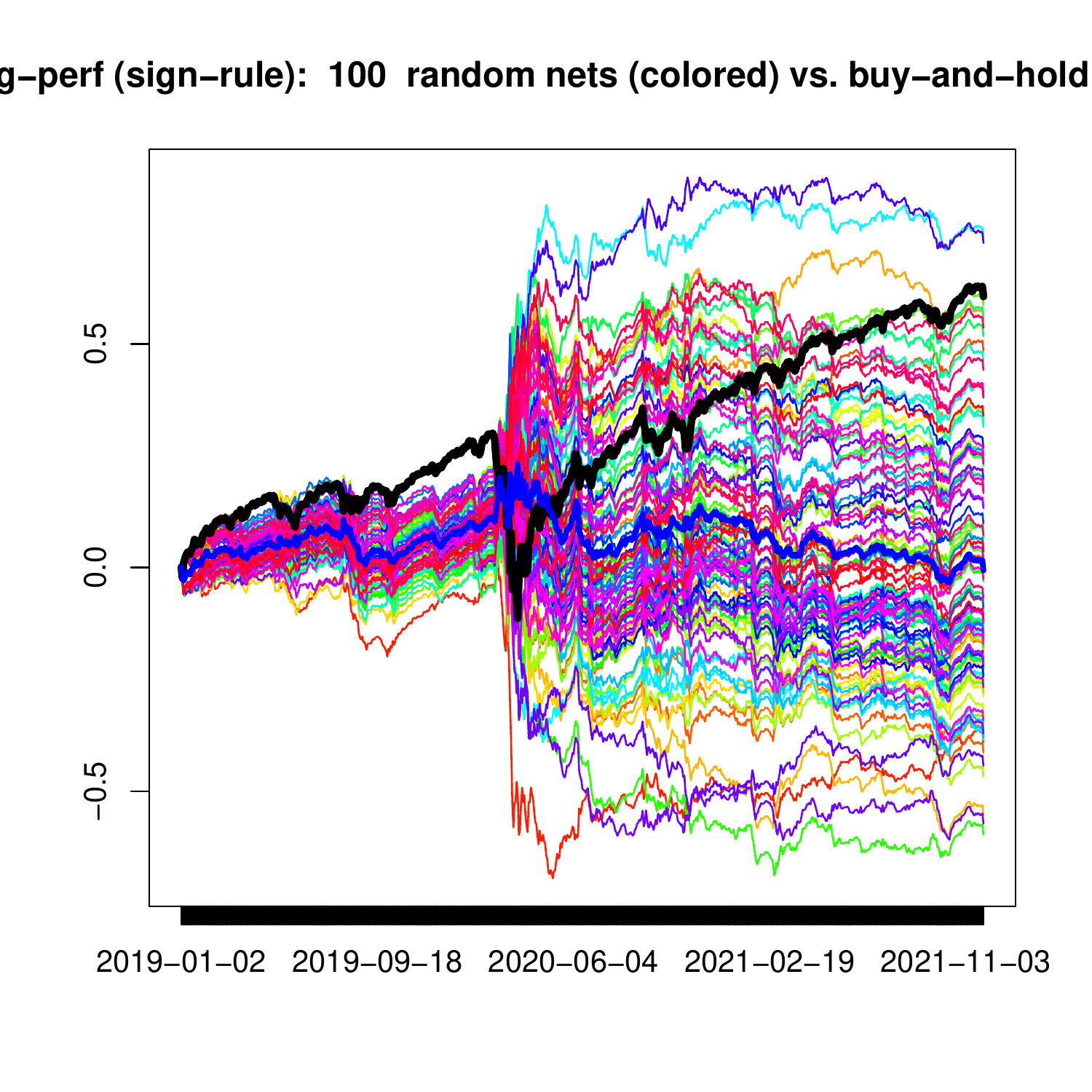}\caption{Out-of-sample trading performances of 100 random nets vs. buy-and-hold (black line)\label{sp500daily_perf_sign_random}}\end{center}\end{figure}

We rely on the previous BTC-framework and apply a feedforward net with a single hidden layer of size 100 to the log-returns of the equity-index. The input layer consists of the last five returns which were identified by classic time series analysis. The in-sample period spans from 2010 up to December 2018 so that the out-of-sample span is subject to severe turmoils, see fig.\ref{sp500daily_perf_sign_random}.
In contrast to the previous BTC-example, the random-nets do not appear to systematically outperform the buy-and-hold benchmark, quite the contrary; moreover, fears around the Covid-outbreak lead to a sudden spread suggesting that the nets are essentially uninformative about the resulting market burst. It is therefore interesting to verify the potential added-value of second-order information in a context where ordinary net-forecasts fail to generate value. Before, though, we briefly address explainability and compare our findings with results obtained for BTC. 




\subsubsection{LPD: Prices vs. Log-Returns}

\begin{figure}[H]\begin{center}\includegraphics[height=3in, width=3in]{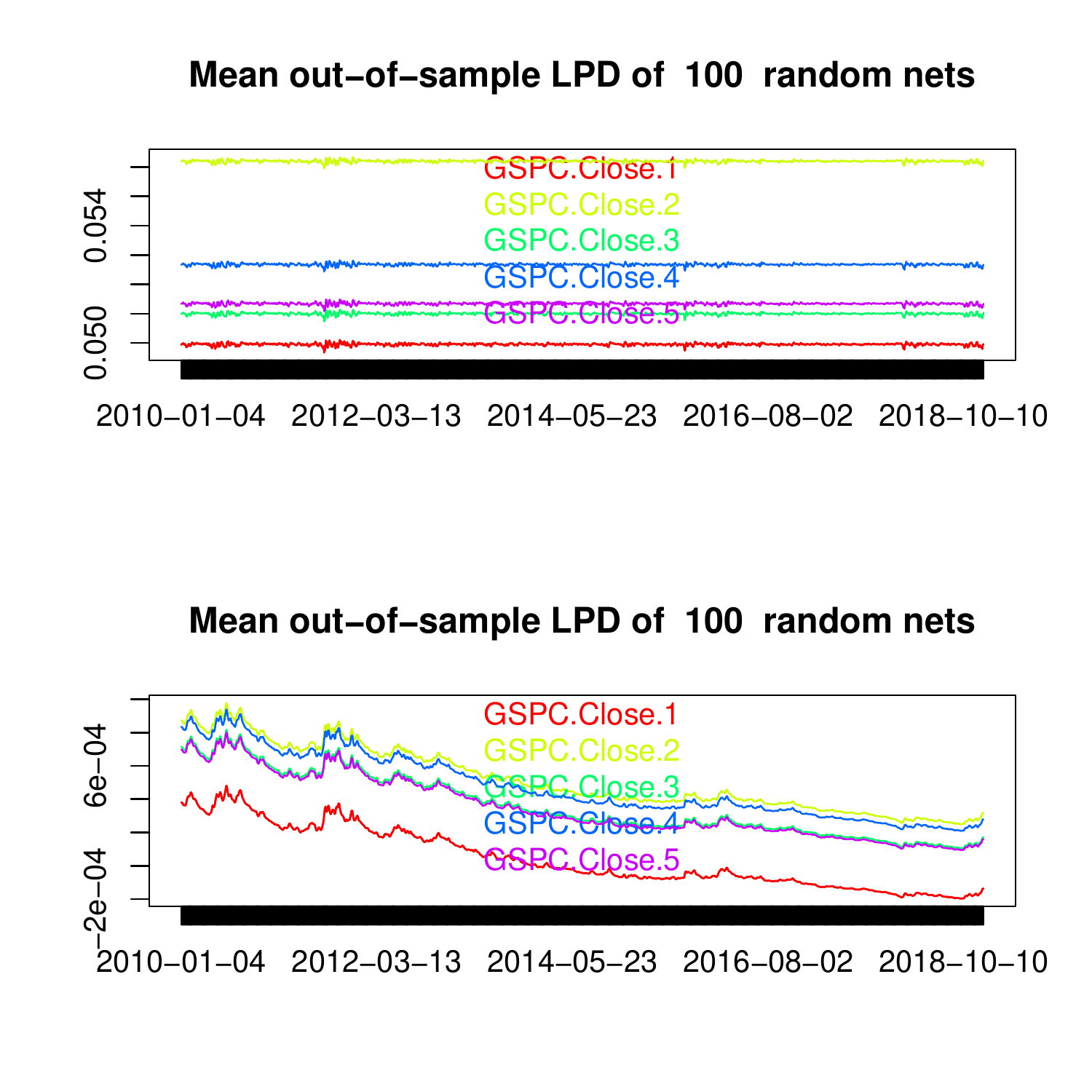}\caption{LPDs of returns (top) and of prices (bottom)\label{sp_daily_LPD_array_out_sample_agg}}\end{center}\end{figure}

We here emphasize the aggregate mean LPDs of the five explanatory lagged input variables, see fig.\ref{sp_daily_LPD_array_out_sample_agg} which compares LPDs of nets fitted to returns (top panel) and of nets fitted to prices (bottom panel).
While the former essentially confirm previous findings, obtained for the BTC in terms of simplistic linear forecast heuristic, the latter suggest a non-stationary pattern of net sensitivities which would be difficult to reconcile with a standard random-walk hypothesis of financial market (price-)data: the un-interpretable sensitivities should trigger doubt or distrust in the net-outputs in this example and the outcome should at least ask for further clarification. In any case, series with a strong trend pose an additional difficulty to nets equipped with bounded activation functions due to potential neuron-saturation so that we may dissuade from applications to price-data in this context. We therefore pursue our analysis in our standard setting based on log-returns.

\subsubsection{Risk-Management: Exploiting Second Order Information of the LPD}

\begin{figure}[H]\begin{center}\includegraphics[height=3in, width=3in]{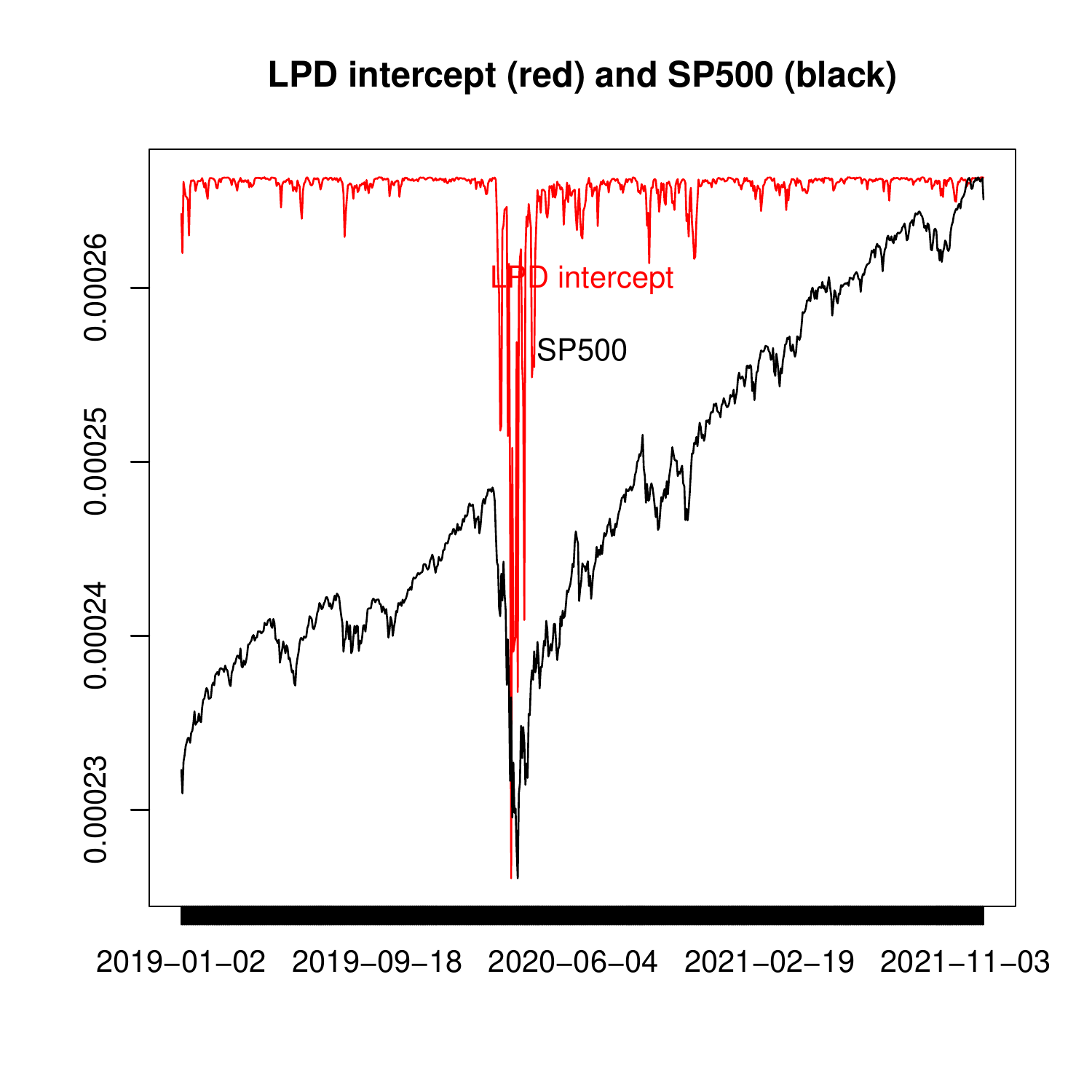}\caption{Out-of-sample LPD of intercept (red) and transformed, scaled and shifted, SP500 (black)\label{sp_daily_LPD_intercept_out_sample_agg}}\end{center}\end{figure}

\begin{figure}[H]\begin{center}\includegraphics[height=3in, width=3in]{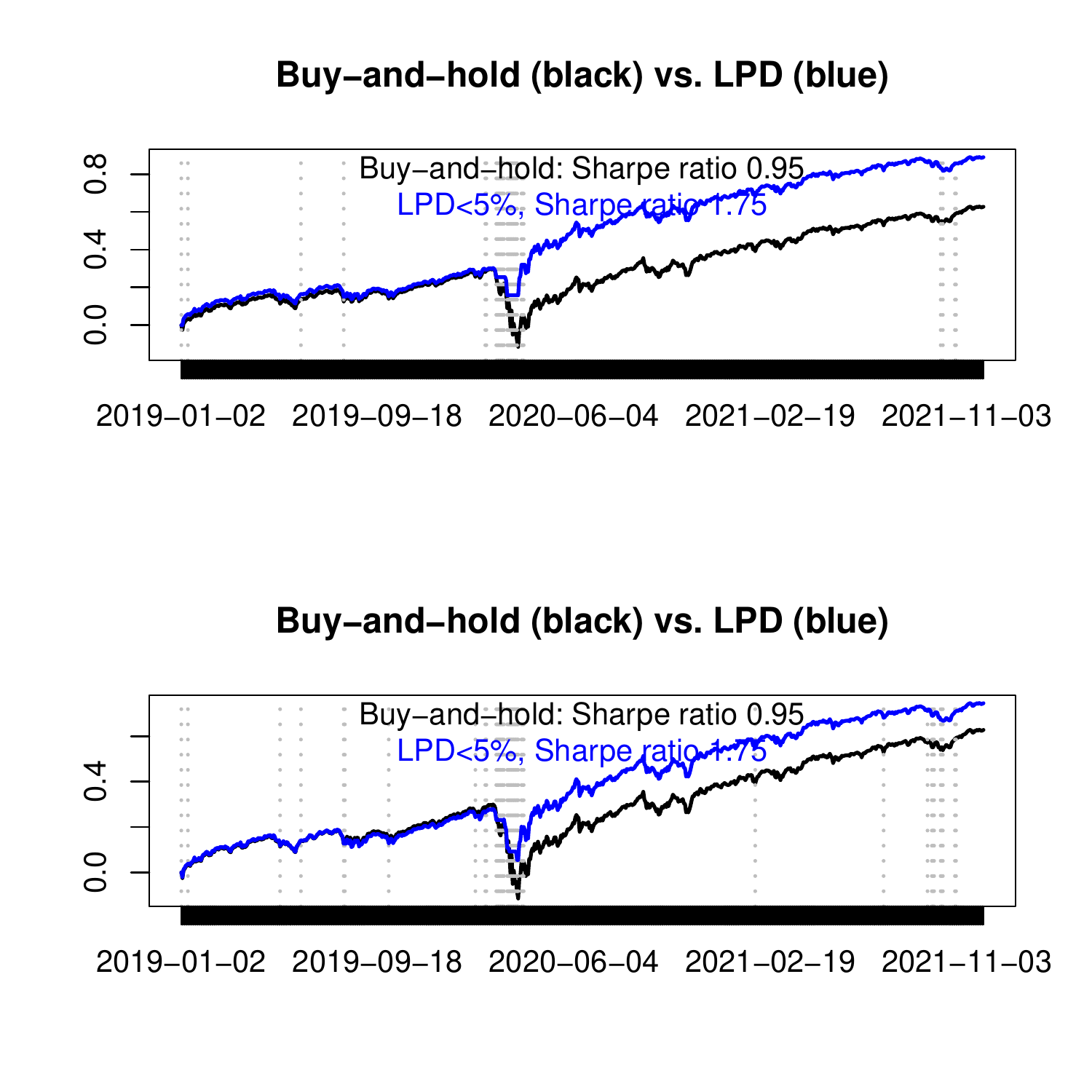}\caption{Buy-and-hold (black) vs. Out-of-sample mean-LPD market-exit strategy: exits (shaded in grey) occur if today's out-of-sample mean-LPD (mean over 100 nets) drops below the 5-percent quantile based on a rolling-window of length one year of its own history. The upper panel is based on the LPD of the intercept alone while the lower panel is obtained from the aggregate LPD of all five explanatory variables plus the intercept.\label{sp_LPD_daily}}\end{center}\end{figure}

As for the previous BTC-example, table \ref{cor_LPD_SP} confirms that second-order non-linear deviations about the mean LPD-level are strongly correlated across explanatory variables (first row) as well as across random-nets (second row): once again, second-order non-linearity appears as a common factor pervading the LPD in all dimensions, thus hinting at the common data and the 'state' of the underlying data-generating process.
\begin{table}[ht]
\centering
\begin{tabular}{rrrrrr}
  \hline
 & Lag 1 & Lag 2 & Lag 3 & Lag 4 & Lag 5 \\ 
  \hline
Correlation across mean LPDs & 1.000 & 0.971 & 0.982 & 0.973 & 0.974 \\ 
  Correlation of mean and random LPDs & 0.960 & 0.974 & 0.976 & 0.977 & 0.976 \\ 
   \hline
\end{tabular}
\caption{Cross-correlations of mean LPDs (first row: all correlations are referenced against the lag-one input variable) and of random LPDs (second row). For each explanatory variable the mean of the 100 random-correlations are reported in the second row.} 
\label{cor_LPD_SP}
\end{table}In contrast to the previous BTC-example, which emphasized a rather short-term perspective, due to frequent local  bursts of the crypto-currency, the S$\&$P equity-index tracks more closely the 'real' economy with longer episodes of economic expansion and regular growth, interrupted by protracted crises. We therefore propose an alternative concept for addressing RM in this context, 
by aiming at a so-called \emph{crisis-triggering}, staying in a long-position most of the time except at severe turmoils, ideally. 
Moreover, we complement the LPD of the input variables with the synthetic intercept \ref{intercept} which is a natural candidate for tracking down-turns in terms of unusually small drifts, see fig.\ref{sp_daily_LPD_intercept_out_sample_agg}. With regards to the crisis-triggering, we postulate that the severity of a crisis can be expressed in terms of mean duration and frequency: we here assume half-a-year for the duration and once-per-decade for the frequency, to obtain a corresponding $0.5/10=5\%$-quantile for the LPD. Furthermore, assuming a sample of size ten for determining the $5\%$-quantile we can select a rolling window of length $\frac{10}{0.05}=200$ i.e. roughly a year.
The performance of the resulting 'lazy' active strategy, relying on the intercept alone, is displayed in the upper panel of fig.\ref{sp_LPD_daily}. The lower panel is based on a combination of all $5+1=6$ LPDs whereby the market exposure is sized according to the simple aggregate rule
\begin{eqnarray}\label{agg_lpd}
\frac{1}{6} \sum_{i=1}^6 I_{\{LPD_{it}>5\%-quantile\}}
\end{eqnarray}
where the indicator function $I_{\{LPD_{it}>5\%-quantile\}}$ is one or zero depending on the $i$-th LPD being above its $5\%$-quantile at time $t$, or not.
As intended, exit signals are scarcely distributed and mainly concentrated towards episodes of sustained uncertainty and accordingly more or less protracted down-turns. Although the LPD corresponding to the intercept appears to be endowed with enhanced timing-ability in this example, the aggregate LPD defined in \ref{agg_lpd} might end-up as a more convincing instrument for the monitoring and management of crises because it supports also a measure of the dependence-structure of the data. Finally, the example illustrates that the LPDs of the nets can provide added-value despite subpar forecast and hence trading performances of the latter, as found in fig.\ref{sp500daily_perf_sign_random}.

\section{Conclusion} \label{conclusion}

The need for opening the "black box" of machine learning has gained great traction in the past decade, as the need for controlling these models and regulatory concerns, have increased. Solutions to this issue fall within the so-called explainable AI field which aims at producing methods that enable users to understand and appropriately trust outputs generated from AI-based systems. Although the literature is offering an ever-growing suite of such XAI techniques, research on XAI methods specifically suited for financial time series remains limited. Furthermore, classical XAI approaches and their implementations cannot be easily adjusted to correctly account for the time dependency of financial data which in turn makes their application to this domain very limited.\\
For the purpose of addressing this gap, we propose a time-series approach to explainability which preserves dependency-structures by emphasizing infinitesimal changes of the explanatory variables on some X-function of the neural net's output. 
We propose a family of vanilla and customized X-functions addressing various explainability-prospects, including linear replication (LPD), departures from linearity (QPD), overfitting (IPD) and customized functions. We also provide formal derivations of net sensitivities for generic differentiable X-functions. 
Our empirical examples, based an applying the LPD to financial data, suggest evidence of  unexpectedly simple net structures, replicating well-known forecast heuristics.  However, on top of the somehow crude first order approximation, our examples also highlight the existence of a weak but pervasive process, a second-order non-linearity factor common to random LPD-realizations as well as to explanatory variables, which tracks responses of the nets to changes in the data generating process. An application of simple diagnostic tools to the extracted factor can help identify singular events or episodes likely to affect normal operation mode (fraud) or risk-perception (risk-management). In this sense, we argue that our XAI-tool can contribute to a better understanding of a phenomenon by a better explanation of its modelling.


\bibliographystyle{unsrt}
\bibliography{references}


\end{document}